\documentclass[11pt]{article}  
\usepackage{amsmath,amssymb,epsfig,array,calc,multirow,amsfonts}

\def\p{{\mathbf p}}

\def\mpc{h^{-1}{\rm Mpc}}
\def\x{{\mathbf x}}
\def\V{{\mathbf V}}

\def\tv{{\mathbf t}}
\def\Kc{{\mathbb K}}
\def\E{{\mathbb E}}
\def\Po{{\mathbb P}}
\def\phiv{{\vec{\varphi}}}
\def\cv{{\mathbf c}}
\def\phiv{{\vec{\varphi}}}

\def\K{{\mathcal K}}
\def\d{{\rm d}}
\def\CD{{\cal D}}
\def\dr{{\rm d}}
\newcommand{\daverage}[1]{\left\langle #1 \right\rangle_{\rm u}}
\begin{document}
\normalsize
\newcounter{cureqno}%
\newenvironment{mathletters}{%
 \refstepcounter{equation}%
 \setcounter{cureqno}{\value{equation}}%
 \edef\@tempa{\theequation}%
 \expandafter\def
 \expandafter\theequation
 \expandafter{\@tempa\alph{equation}}%
 \setcounter{equation}{0}%
}{%
 \let\theequation\
 \setcounter{equation}{\value{cureqno}}%
}%
\newcommand\eqnum[1]{%
 \def\theequation{#1}%
 \let\@currentlabel\theequation
 \addtocounter{equation}{\m@ne}%
}%
\centerline{\Large{\bf Morphometry of Spatial Patterns}}

\bigskip\medskip

\centerline{Claus Beisbart$^1$, Thomas Buchert$^{2}$ and Herbert Wagner$^1$}
\centerline{\bf 7/2000}

\bigskip\bigskip
\centerline{$^1$Theoretische Physik, Ludwig--Maximilians--Universit\"at}
\smallskip
\centerline{Theresienstr. 37, D--80333 M\"unchen, Germany}
\smallskip
\centerline{$^2$Theoretical Astrophysics Division, National Astronomical Observatory}
\smallskip
\centerline{ 2--21--1 Osawa Mitaka Tokyo 181--8588, Japan }
\smallskip
\normalsize

\vspace{5mm}
\bigskip
\begin{abstract}
 
  Minkowski functionals constitute a  family of order parameters which
  discriminate   spatial  patterns  according   to  size,   shape  and
  connectivity. Here  we point out, that these  scalar descriptors can
  be complemented  by vector--valued curvature measures  also known as
  Querma{\ss}  vectors.    Using  examples  of   galaxy  clusters,  we
  demonstrate   that  the   Querma{\ss}  vectors   provide  additional
  morphological  information on  directional  features and  symmetries
  displayed by spatial data.
\end{abstract}

\centerline{\it Physica A }

\bigskip\medskip

\bigskip\medskip


\bigskip\medskip
\newpage


\section{Introduction}
\label{sec:in}
\begin{figure}
\begin{minipage}[t]{.99\linewidth}
\centering
\epsfig{file=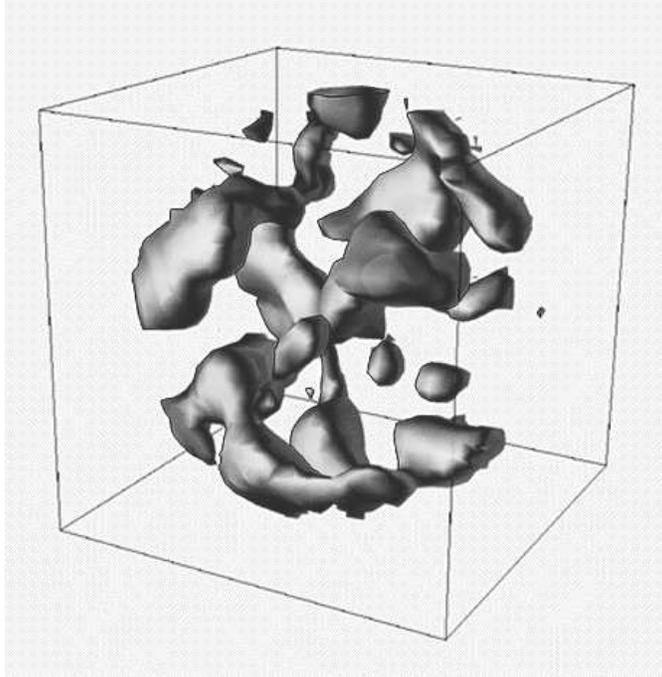,height=9cm}
\end{minipage}\hfill
\caption{ An isodensity contour  of the galaxy distribution within the
PSCz      survey       (from      the      PSCz       homepage      at
http://star-www.dur.ac.uk/cosmology/theory/pscz.html).
\label{fig:gal}}
\end{figure}

Spatial  patterns  originating  from  the polymorphic  aggregation  of
matter  in Nature occur  on vastly  different length--scales  and with
unlimited variety.  The pattern shown in Figure \ref{fig:gal} displays
an  isodensity contour  of  the smoothed  galaxy  distribution in  our
Universe, but it could  also represent a biological structure obtained
with  X--ray tomography, for  instance.  \\  The visual  impression of
such patterns  may already convey  valuable insights on  the formation
processes  or  the functionality  of  those  structures. However,  for
unbiased inferences as well  as for comparison with model simulations,
for  example,  one  needs   objective  and  quantitative  measures  to
characterize the  geometry and topology of  typical structural motifs.
\\  In the present  paper we  discuss a  family of  suitable measures,
which  include,  besides   the  scalar  Minkowski  functionals,  their
vectorial  relatives  --   the  curvature  centroids,  or  Querma{\ss}
vectors.  Taken  together, these measures constitute  a complete class
of morphometric  order--parameters for  spatial patterns formed  -- or
approximated -- by the set  union and intersection of an arbitrary but
finite number  of convex ``pixels'' or ``voxels''.   \\ The usefulness
of Minkowski functionals has already been demonstrated by the analysis
of    galaxy    catalogues    and    simulations    of    large--scale
structure~\cite{kerscher:abell,kerscher:fluctuations,sahni:shapefinders,schmalzing:webI},
as well  as of the temperature  distribution in maps  of the microwave
background  radiation~\cite{schmalzing:cmb}.   Furthermore, they  were
used     to     characterize     the    morphology     of     spinodal
decomposition~\cite{mecke:spinodal}  and  of  patterns arising  within
reaction--diffusion  systems~\cite{mecke:reaction}.   Here,  we  focus
attention  on the  less  familiar Querma{\ss}  vectors and  illustrate
their  versatility  with the  examples  of  galaxy  clusters and  with
patterns generated by lattice automata.


\section{Morphometric order parameters}
\label{sec:quer}
In this section we recall the definitions of Minkowski functionals and
Querma{\ss}  vectors. The  presentation will  be brief  and  informal. 
Readers,  who  are interested  in  mathematical  details, may  consult
refs.~\cite{hadwiger:vekt,schneider:schwerpunkte,schneider:schwerpunkte2,hadwiger:vect2,mecke:robust}.
\\
Consider  a   compact  convex  point  set  ({\em   body})
$K\subset\E^d$, in Euclidean d--dimensional space with volume
\begin{equation}
V_0 (K)  = \int_K \dr^d \x > 0
\end{equation}
and with center of mass at the position:
\begin{equation}
\p_0 \equiv \V_0/V_0\quad,\quad \V_0 (K) =  \int_K\x \dr^d \x\;\;\;.
\end{equation}
Now, we  let the body  grow to form  a parallel body  $K_\epsilon$,
$\epsilon > 0$, by including  all points $\x$ within a distance
$d(K,\x)\leq\epsilon$ to $K$. From Steiner's theorem~\cite{mcmullen:valuations} we know  that the volume
$V_0(K_\epsilon)$ of  $K_\epsilon$ is a  finite polynomial in
$\epsilon$,
\begin{equation}
V_0 (K_\epsilon) = \sum_{i=0}^{d} \binom{d}{i} \omega_{i}V_{i}(K) \epsilon^i\;\;\;,
\end{equation}
where $\omega_i= \pi^{i/2}/\Gamma (1 + i/2)$ denotes the volume of the
$i$--dimensional unit ball.  The coefficients $V_i(K)$ define the {\it
Minkowski   functionals}  of   the   body  $K$.    Since  the   volume
$V_0(K_\epsilon)$  depends  on  the  shape  of  $K$,  the  functionals
$V_i(K)$ contain morphological information on the original body.  \\ A
corresponding      expansion     holds      for      the     vectorial
$\V_0(K)$~\cite{hadwiger:vekt},
\begin{equation}
\V_0 (K_\epsilon) = \sum_{i=0}^{d} \binom{d}{i}  \omega_{i} \V_{i}(K) \epsilon^i\;\;\;,
\end{equation}
which defines the {\em Querma{\ss}  vectors} $\V_i(K), \;0\leq i\leq d$. 
Under  translations  (t)  and  rotations  (r) of  the  body  $K$,  the
Minkowski functionals are {\em motion--invariant} scalars, whereas the
Querma{\ss} vectors are {\em motion--equivariant}:
\begin{mathletters}
\begin{eqnarray}
\V_i (t K) =& \V_i (K) + \tv V_i(K),      \\ 
\V_i (r K) = &R(r)\V_i (K),              
\end{eqnarray}
\end{mathletters}
\setcounter{equation}{5} 
\hspace{-5.5mm} where  $R(r)$ denotes an  orthogonal rotation
matrix and $\tv$ a translation  vector. Both the scalars $V_i$ and the
vectors $\V_i$ are {\em  additive} functionals in the following sense:
If  the  point  set  constituting  the body  $K\subset  \E^d$  is  the
set--union of two bodies $K_1$ and $K_2$, then
\begin{equation}
\phiv_i (K_1 \cup K_2) = \phiv_i (K_1) + \phiv_i (K_2) - \phiv_i(K_1 \cap K_2)\;\;\;,
\label{eq:add}
\end{equation}
\begin{figure}
\begin{minipage}[t]{.99\linewidth}
\centering
\epsfig{file=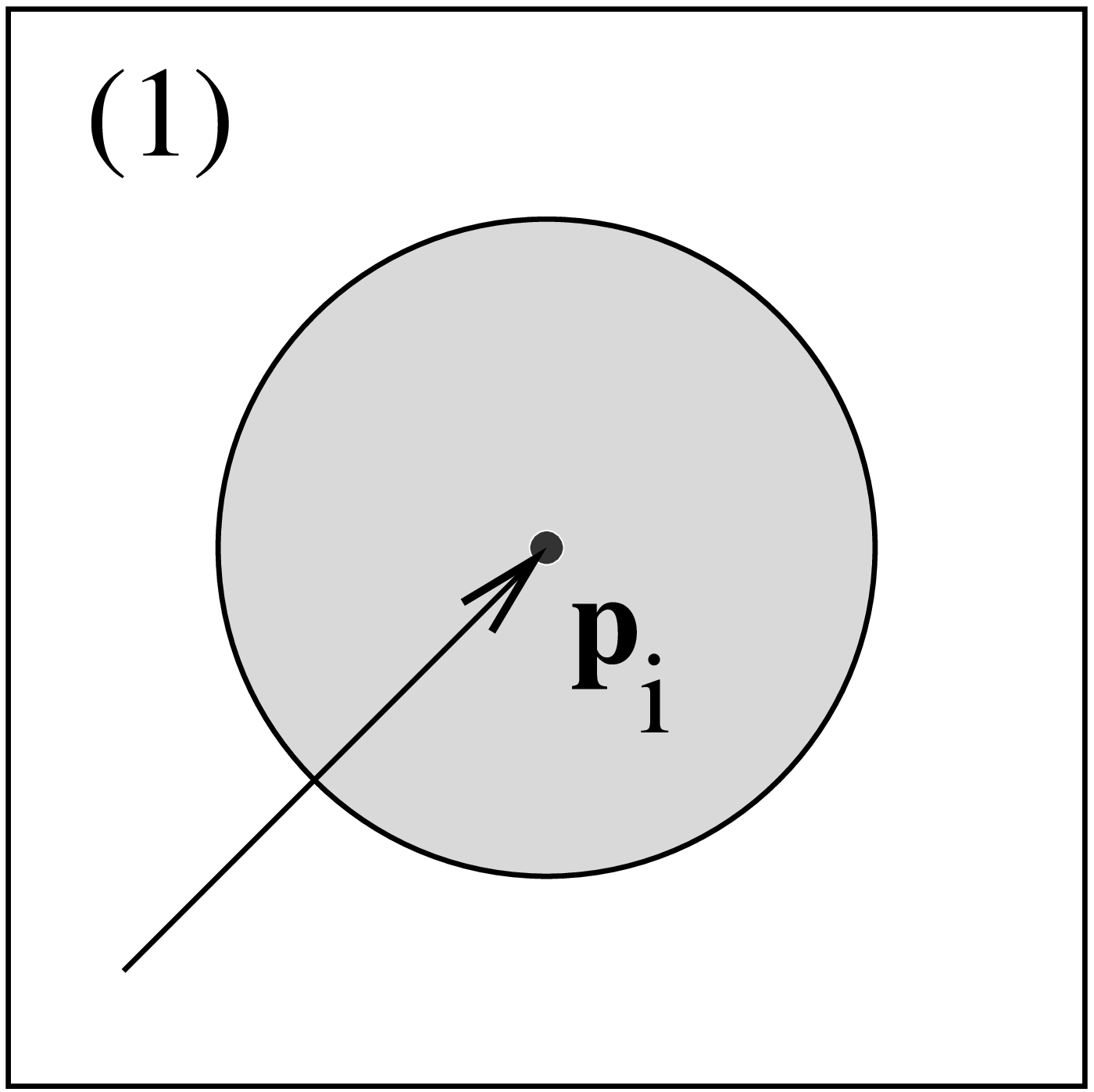,height=4cm}
\epsfig{file=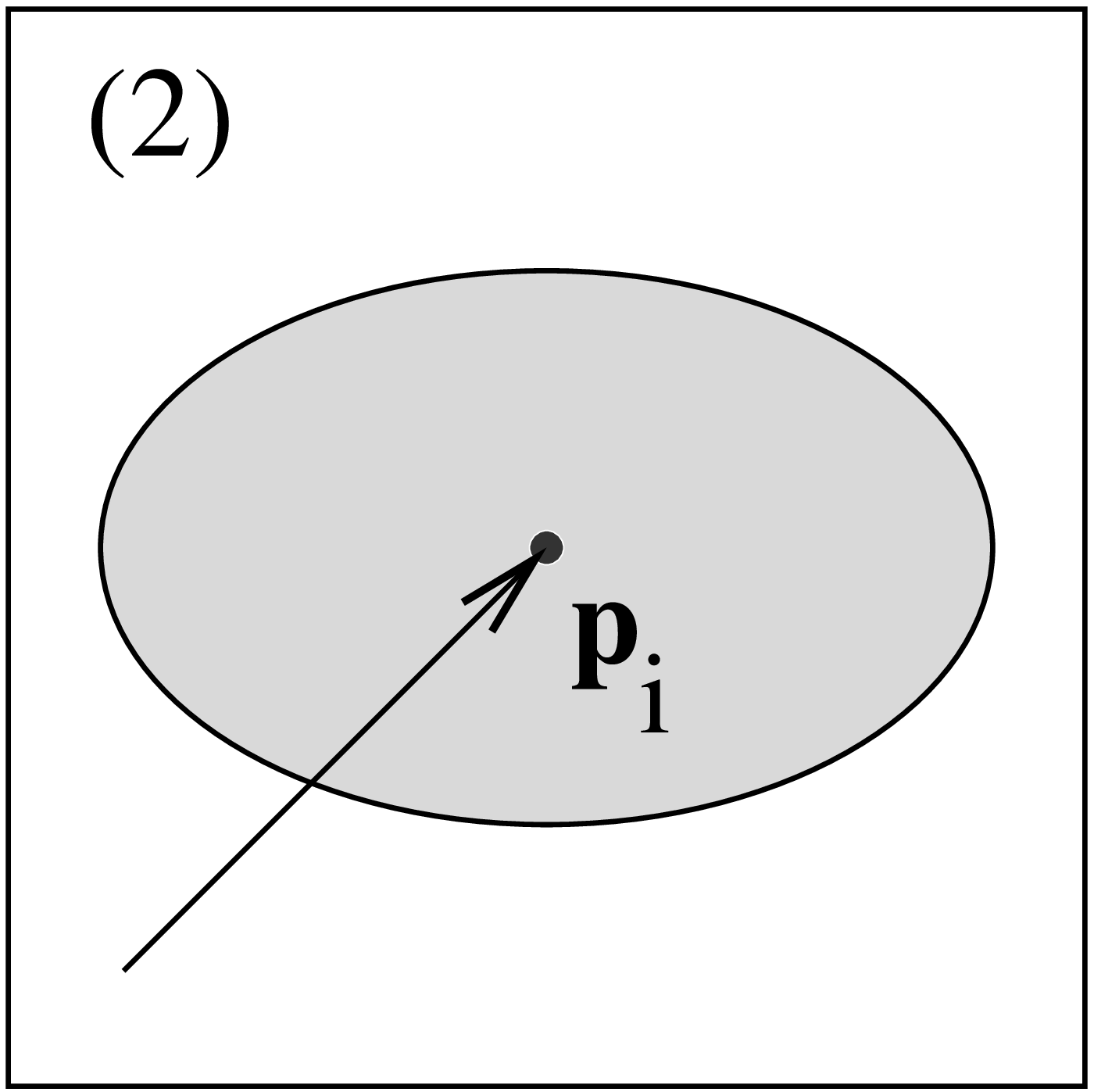,height=4cm}
\epsfig{file=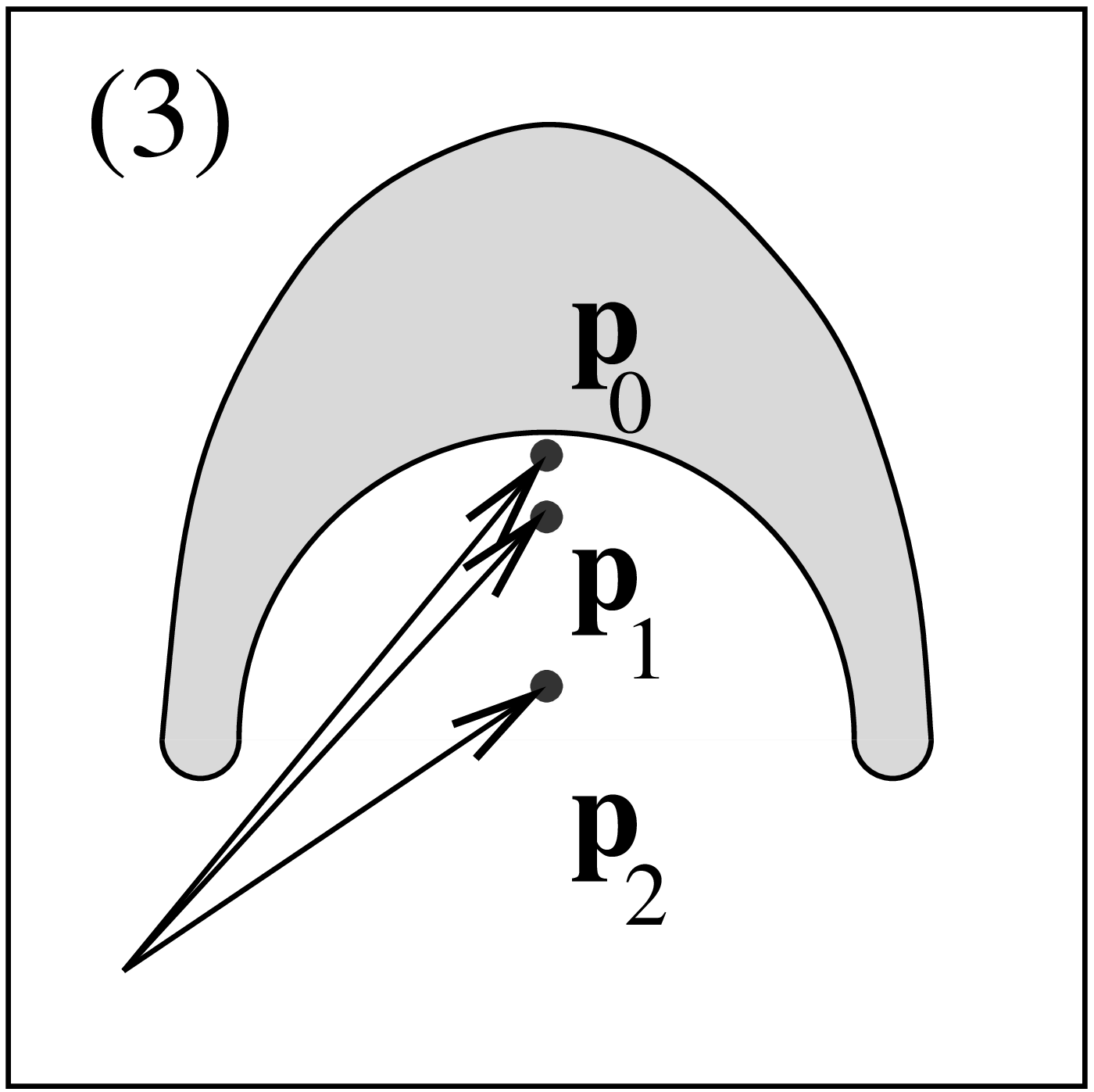,height=4cm}
\epsfig{file=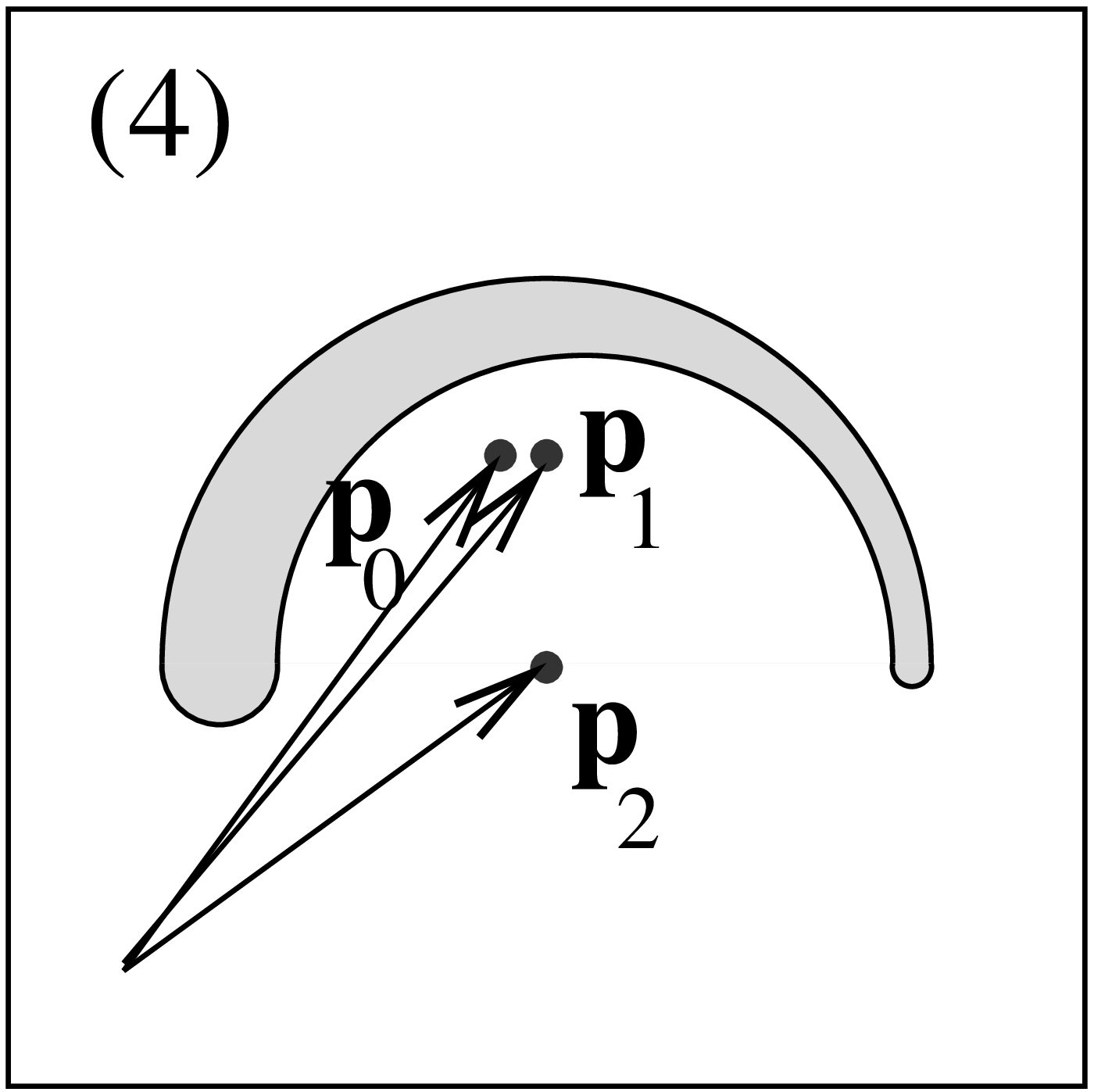,height=4cm}
\epsfig{file=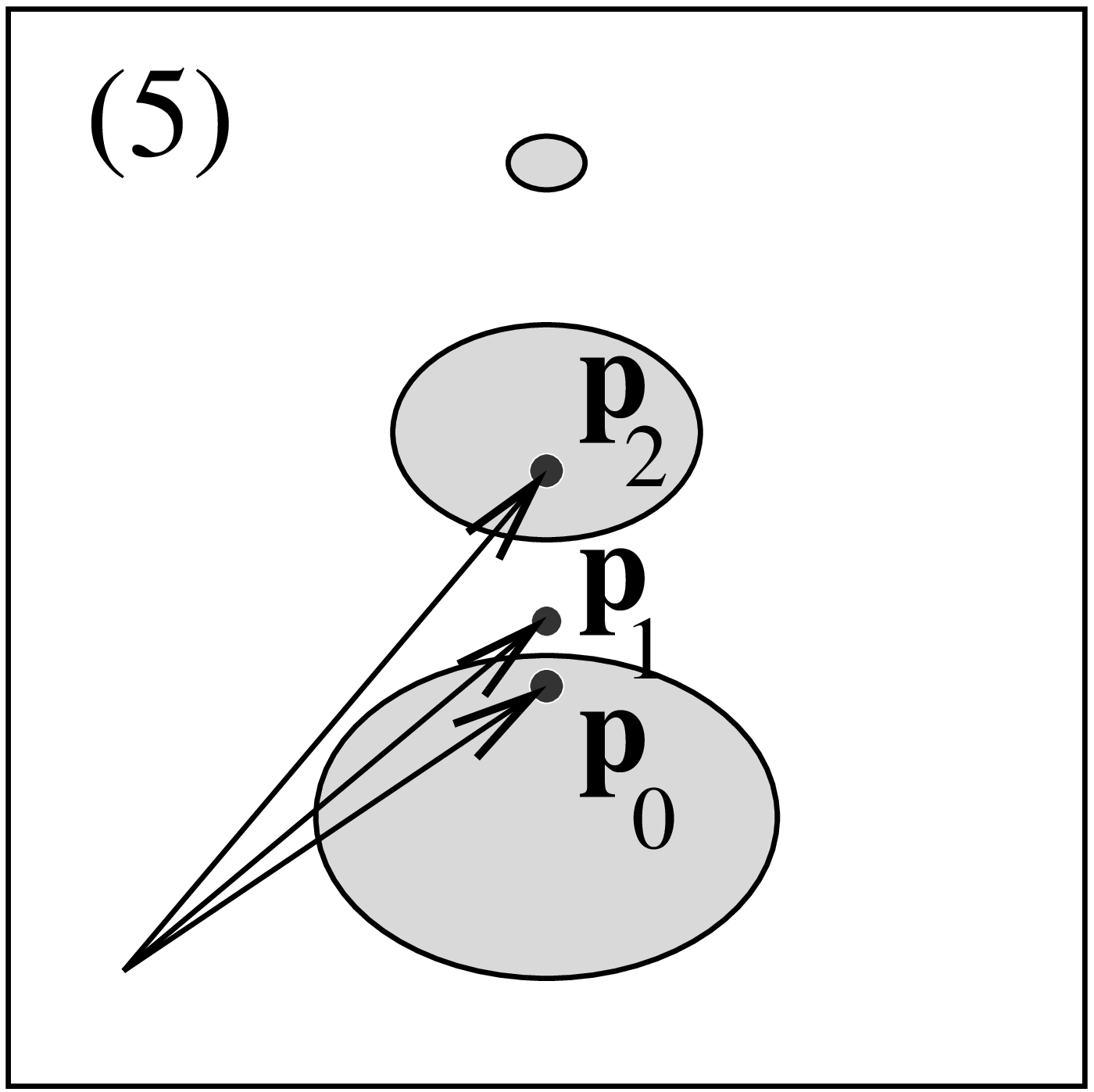,height=4cm}
\epsfig{file=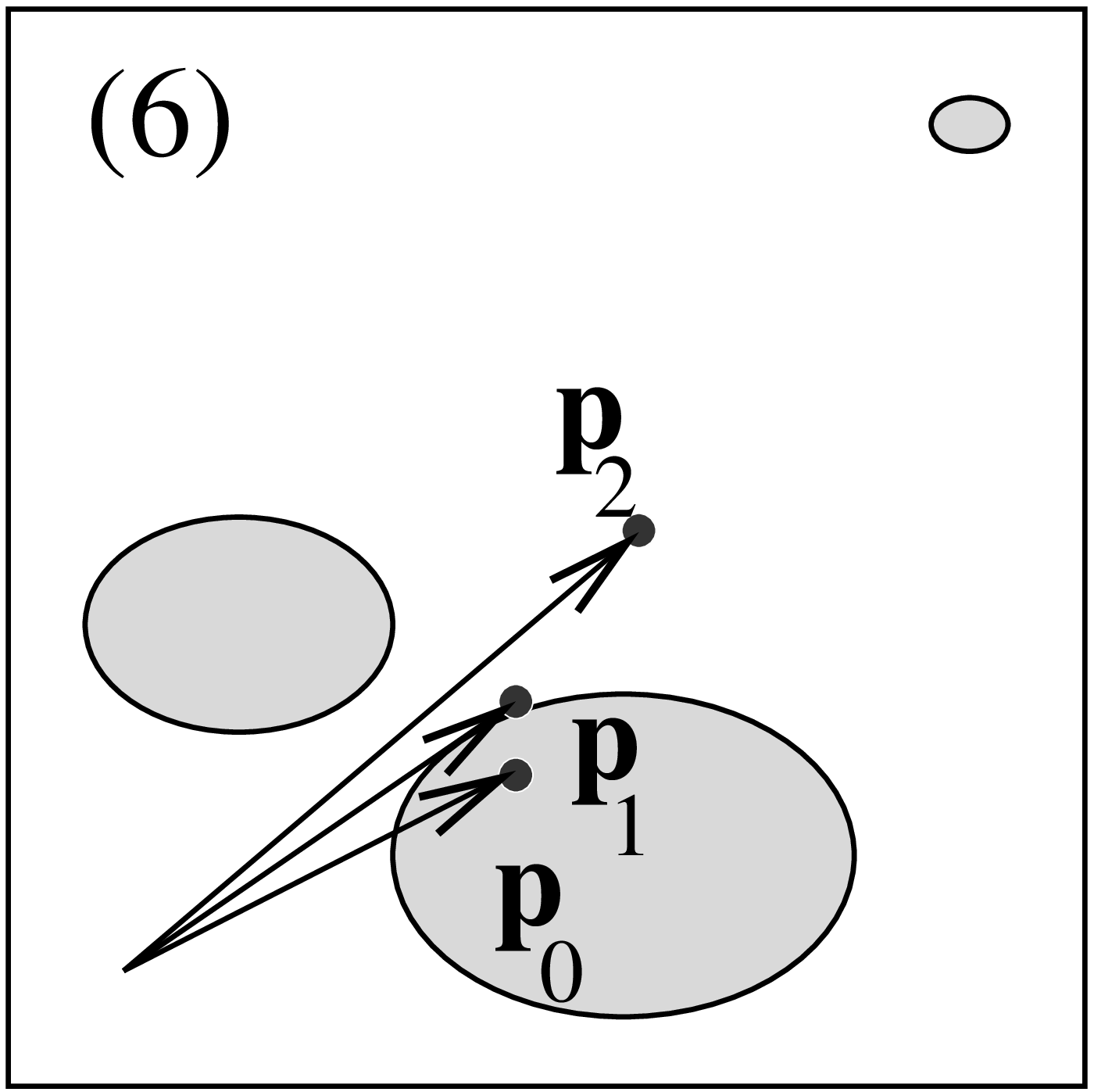,height=4cm}
\end{minipage}\hfill
\caption{Examples of  centroids for a series of  simple patterns.  The
panels (1)--(4) show how  the curvature centroids reflect the symmetry
of a ``cluster''.  For patterns  consisting of more than one body, the
centroids weight  the components  according to their  scalar Minkowski
functionals.  Note that  the  patterns in  panel  (5) and  (6) can  be
discriminated  using the centroids,  but not  by merely  employing the
scalar Minkowski functionals.\label{fig:clusters}}
\end{figure}
where  we use  the compact  notation $\phiv  \equiv  \{V_i,\V_i\}$. If
$K_1$ and  $K_2$ are arbitrary  compact convex bodies, then  the union
set  $K_1\cup  K_2$  is  generally  no  longer  convex.  However,  the
right--hand--side  of Equation  \eqref{eq:add} is  still well--defined
and   may  be   taken   as  the   definition   of  $\phiv_i   (K_1\cup
K_2)$. Consequently, additivity allows us to extend the application of
the  functionals $\phiv$  to patterns  formed by  an  arbitrary finite
union of  convex bodies via iteration of  Equation \eqref{eq:add}.  \\
The functionals $\phiv_i$, for $1\leq i\leq d$, depend on the shape of
the body $K$; if its  surface, $\partial K$, is smooth, with principal
curvatures $c_j  (\x)$, $1\leq j\leq d-1$, then  these functionals may
be expressed  in terms of  surface integrals. For instance,  in $d=3$,
with  the  surface area  element  $\dr  S(\x)$,  and $\vec{z}  \equiv
(1,\x)$ we have:
\begin{gather}
\phiv_1 (K) = \frac{1}{6} \int_{\partial K}  \vec{z} \dr S(\x) \;\;\;,\;\;\;  \phiv_2 (K) = \frac{1}{3\pi} \int_{\partial K}  \frac{1}{2} (c_1(\x) +c_2(\x)) \vec{z} \dr S(\x)\;\;\;, \\
\phiv_3 (K) = \frac{1}{4\pi} \int_{\partial K}  c_1(\x)c_2(\x)  \vec{z} \dr S(\x) \nonumber\;\;\;.
\end{gather}
The vector $\V_3  (K)$ is known as the Steiner point  of the body $K$.
-- We list the  meanings of the Minkowski functionals and  the corresponding vectors in
two  dimensions  in   Table~\ref{tab:meaning}.\\  The  most  prominent
feature  of the  family $\{  \phiv_i( \cdot  ) |  0\leq i\leq  d\}$ is
expressed  by   Hadwiger's  characterization  theorem:   Consider  the
polyconvex ring  $\Po^d$ of spatial patterns, defined  to comprise the
collection  of  arbitrary finite  unions  of  convex  bodies, $\K_N  =
\bigcup_{\alpha=1}^N  K_\alpha \in  \Po^d$, $N<\infty$,  $0  \leq \dim
K_\alpha \leq  d$. According to  the theorem, any additive,  equi-- or
invariant   and   conditionally  continuous\footnote{The   conditional
continuity requires  that a functional  is continuous with  respect to
the Blaschke--Haussdorff metric on  the subset $\Kc^d\subset \Po^d$ of
convex  bodies.}   functional  $\vec{\Phi}$  on  $\Po^d$  is  uniquely
determined by the linear combination:
\begin{equation}
\vec{\Phi} (\K_N) = \sum_{i = 0}^{d} a_i \phiv_i  (\K_N),
\end{equation}
\begin{figure}
\centering
\begin{minipage}[t]{4.7cm}
\epsfig{file=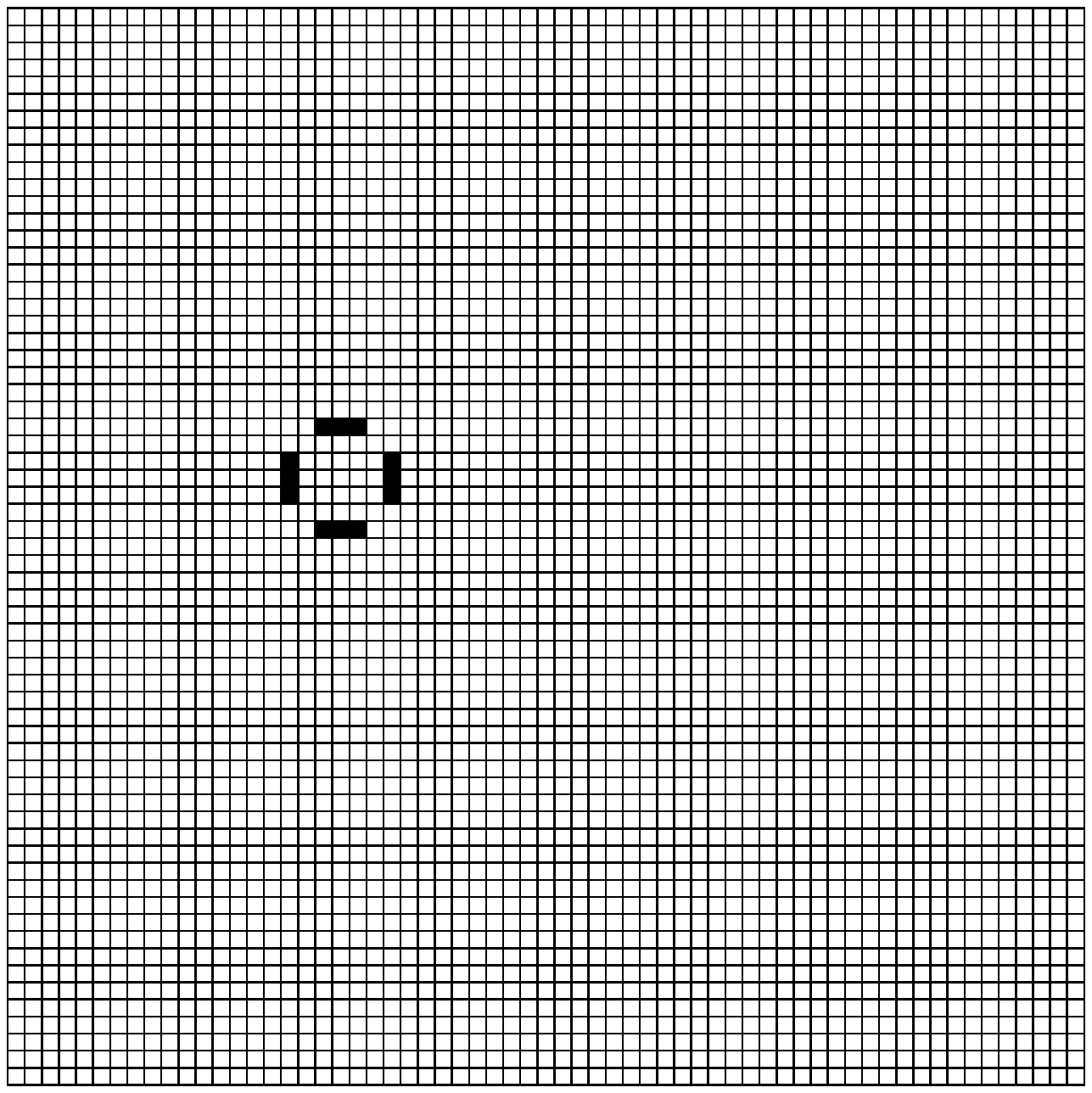,height=4.7cm}
\end{minipage}
\begin{minipage}[t]{0.5cm}
\centering
\vspace{-2.3cm}$\leftarrow$
\end{minipage}
\begin{minipage}[t]{4.7cm}
\epsfig{file=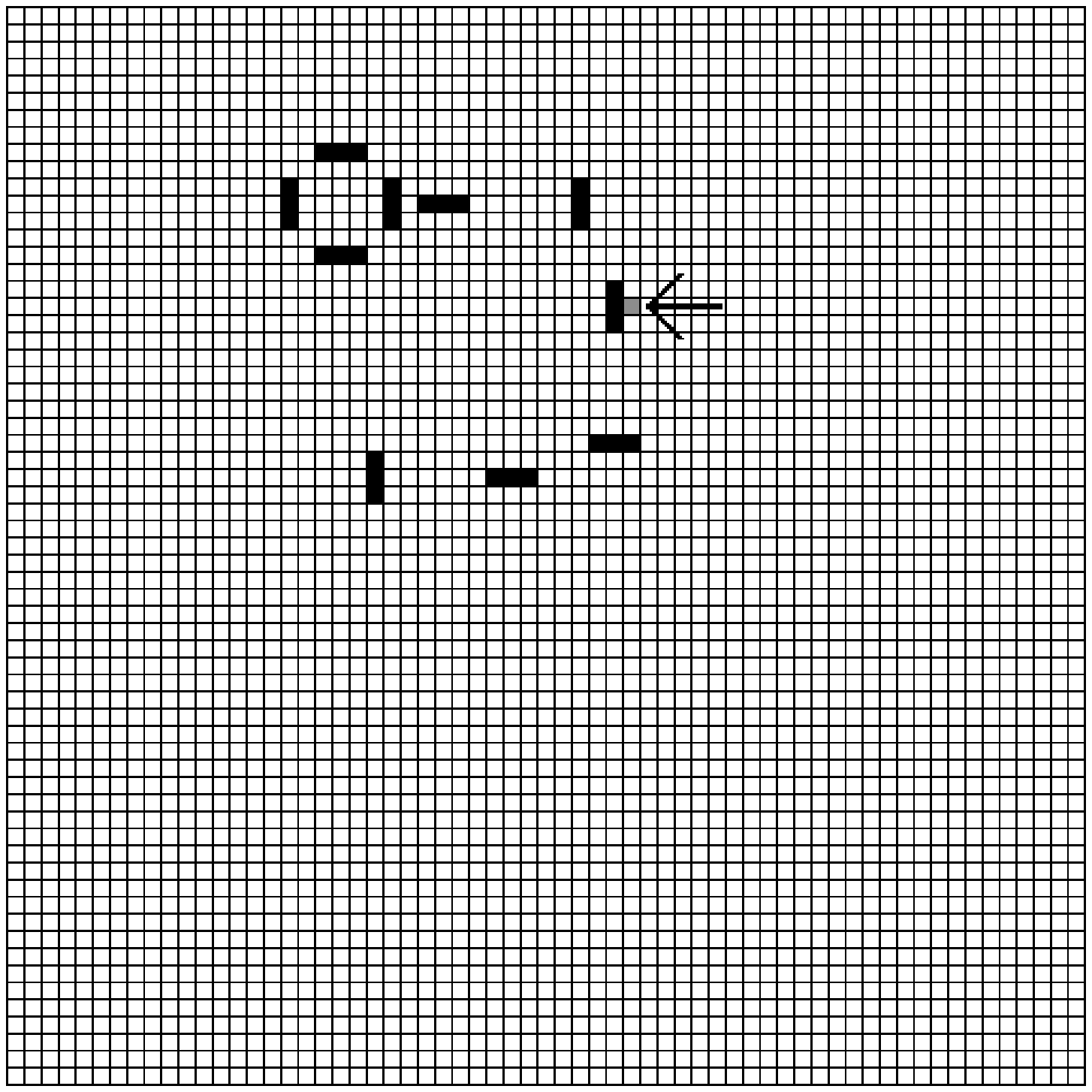,height=4.7cm}
\end{minipage}
\begin{minipage}[t]{0.5cm}
\centering
\vspace{-2.3cm}$\rightarrow$ 
\end{minipage}
\begin{minipage}[t]{4.7cm}
\epsfig{file=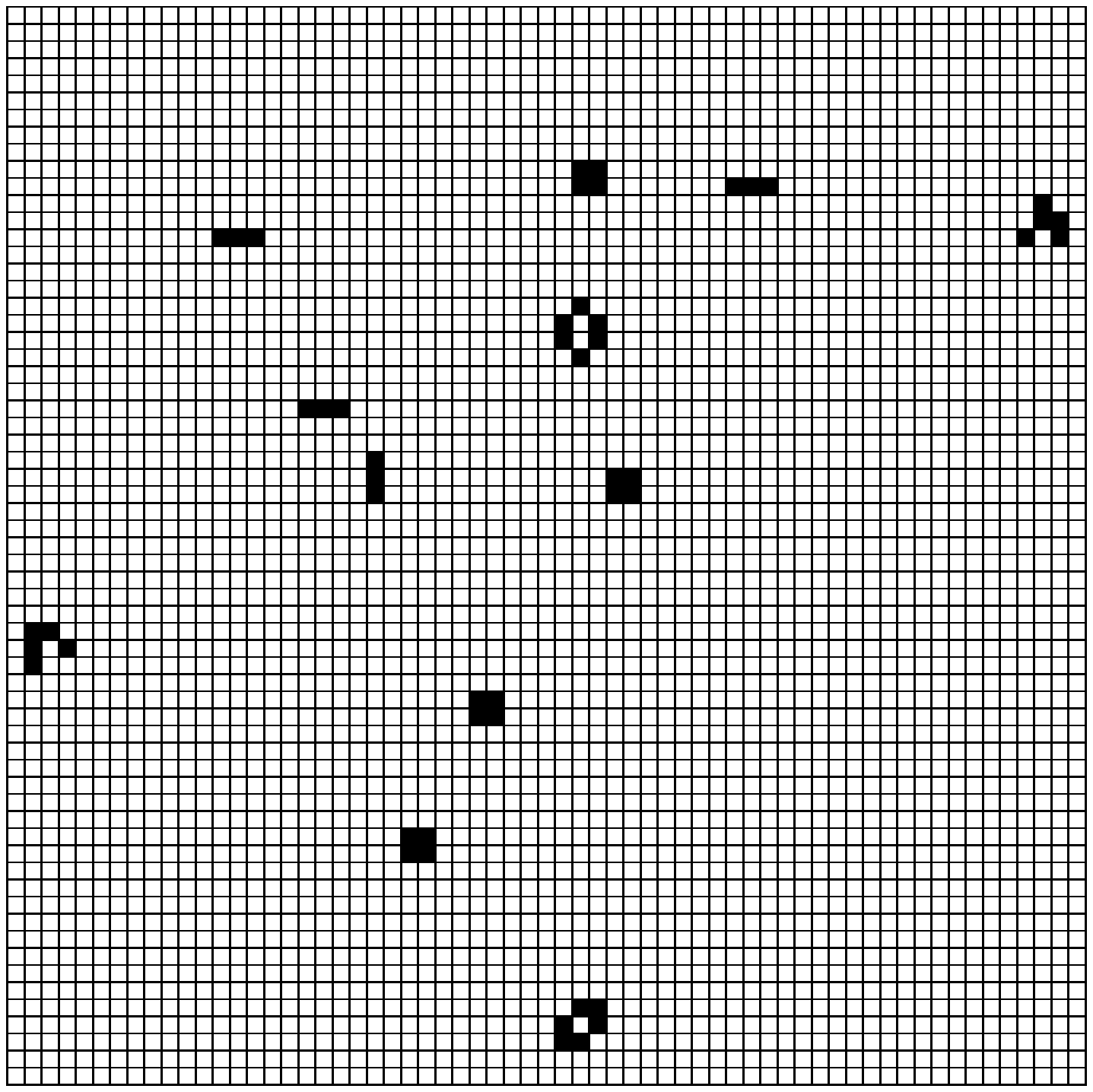,height=4.7cm}
\end{minipage}
\caption{ Two initial configurations  for the ``Game of Life'' (middle
panel)  and the  results  for  the 188th  generation  (right and  left
panel).  In  the initial state, the  ``thunderbirdfuse'' (black cells)
is  perturbed with  the grey  cell marked  by an  arrow.   Whereas the
``thunderbirdfuse''  ends   with  a  ``blinker''   (left  panel),  the
``perturbed  thunderbirdfuse''  evolves into  a  pattern ejecting  two
gliders (right panel).\label{fig:gol}}
\end{figure}
with the real coefficients $a_i$ independent of the pattern $\K_N$. In
this  sense,  the  family  $\{\phiv_i\}$  forms a  complete  basis  of
morphometric descriptors.   \\ For  the practical analysis  of cluster
structures, for  instance, it is convenient to  employ the (curvature)
centroids  $\p_i =\left(  p_i^x,p_i^y,...\right)=  \V_i/V_i$, provided
$V_i\neq  0$,   $i=0,..,d$.   \\  Let   us  consider  the   series  of
2d--clusters shown in Figure \ref{fig:clusters} to gain some intuition
on the kind of information  supplied by the functionals $\phiv_i$, and
the centroids,  in particular.  \\  The clusters (1)--(4)  display the
effect of  symmetry reduction.  In  the case of  the disk (1)  and the
ellipse (2), the centroids  coincide at the symmetry center.  However,
\begin{figure}
\begin{minipage}[t]{.99\linewidth}
\centering
\epsfig{file=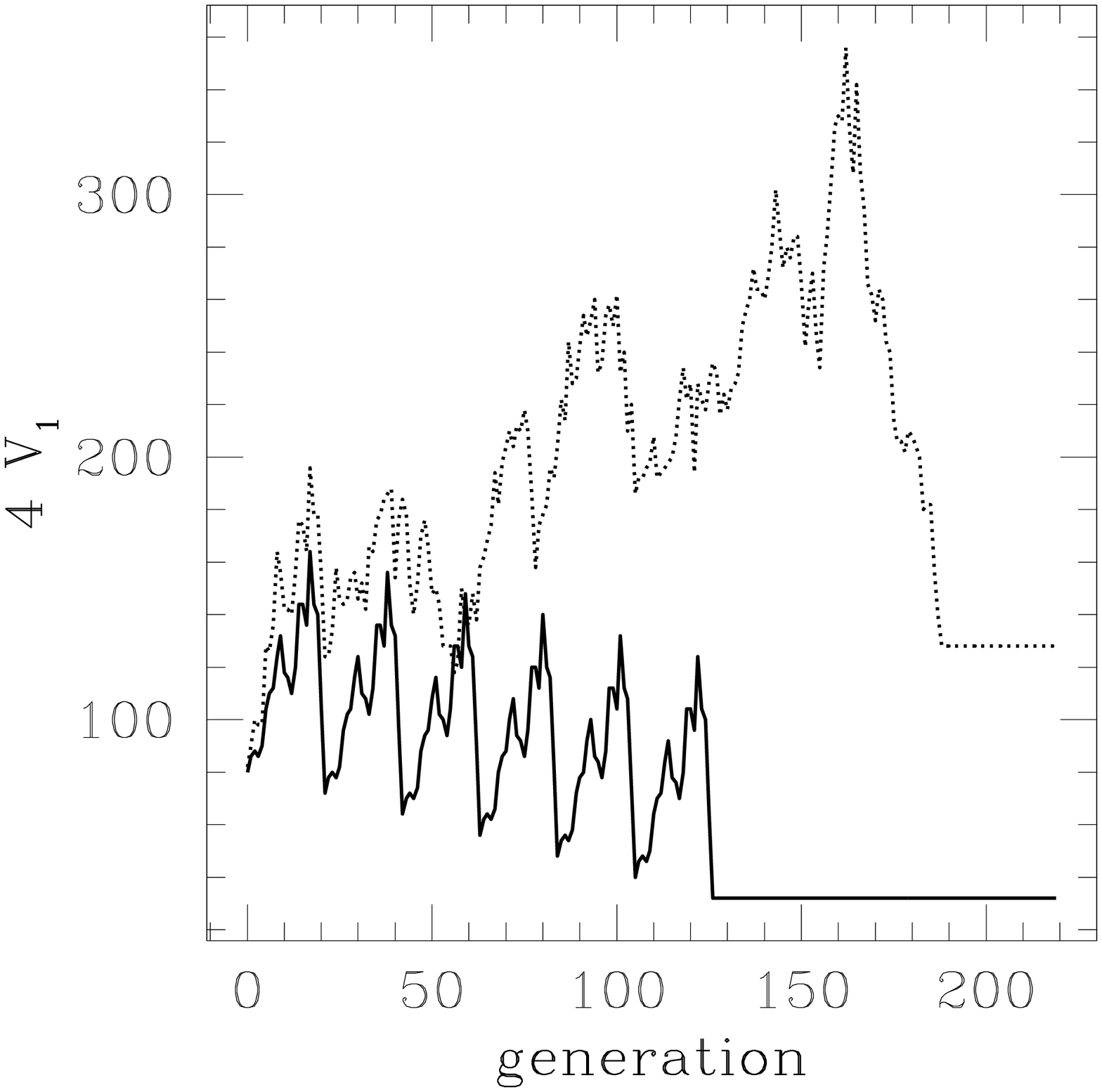,height=5cm}
\epsfig{file=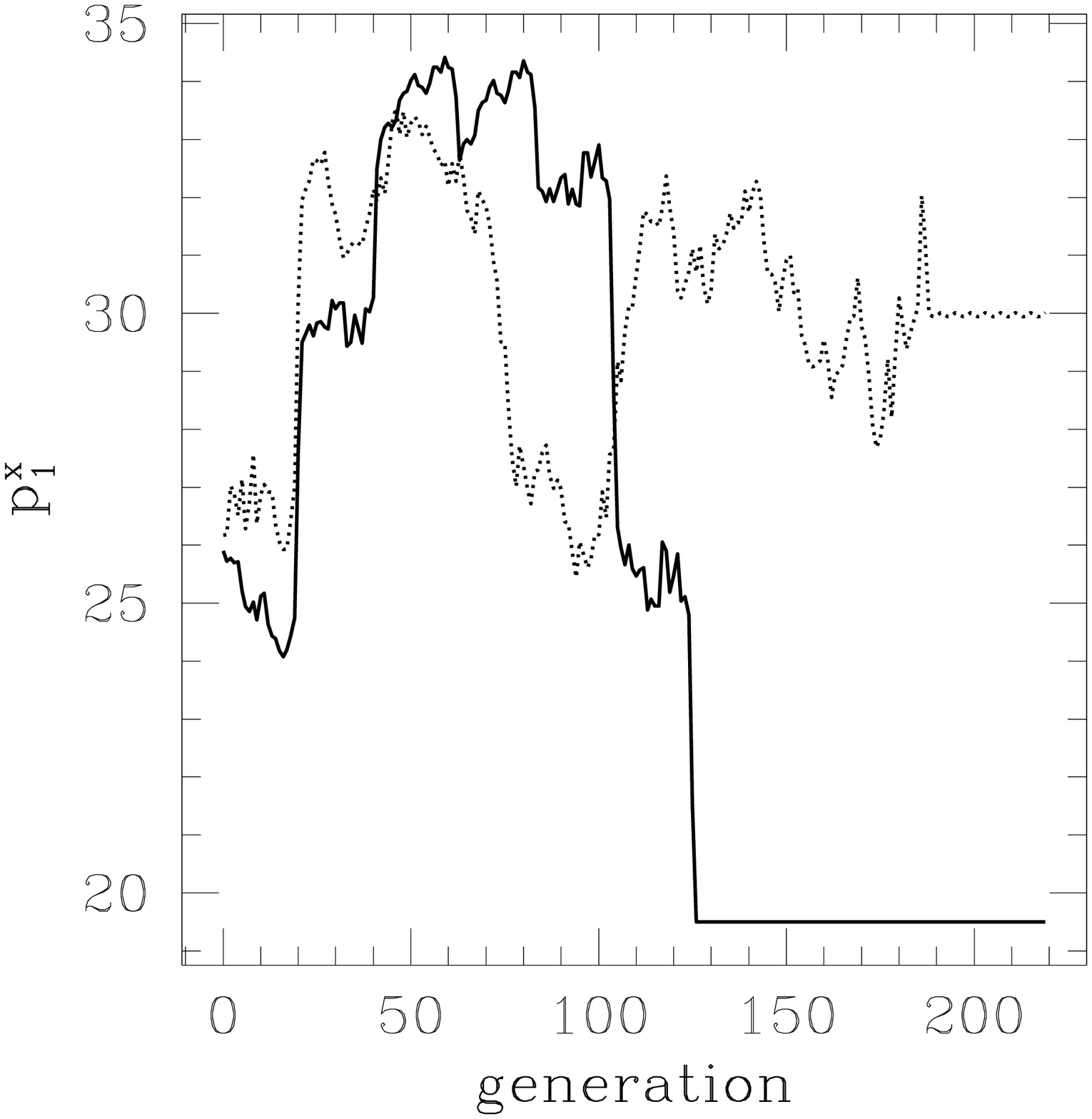,height=5cm}
\epsfig{file=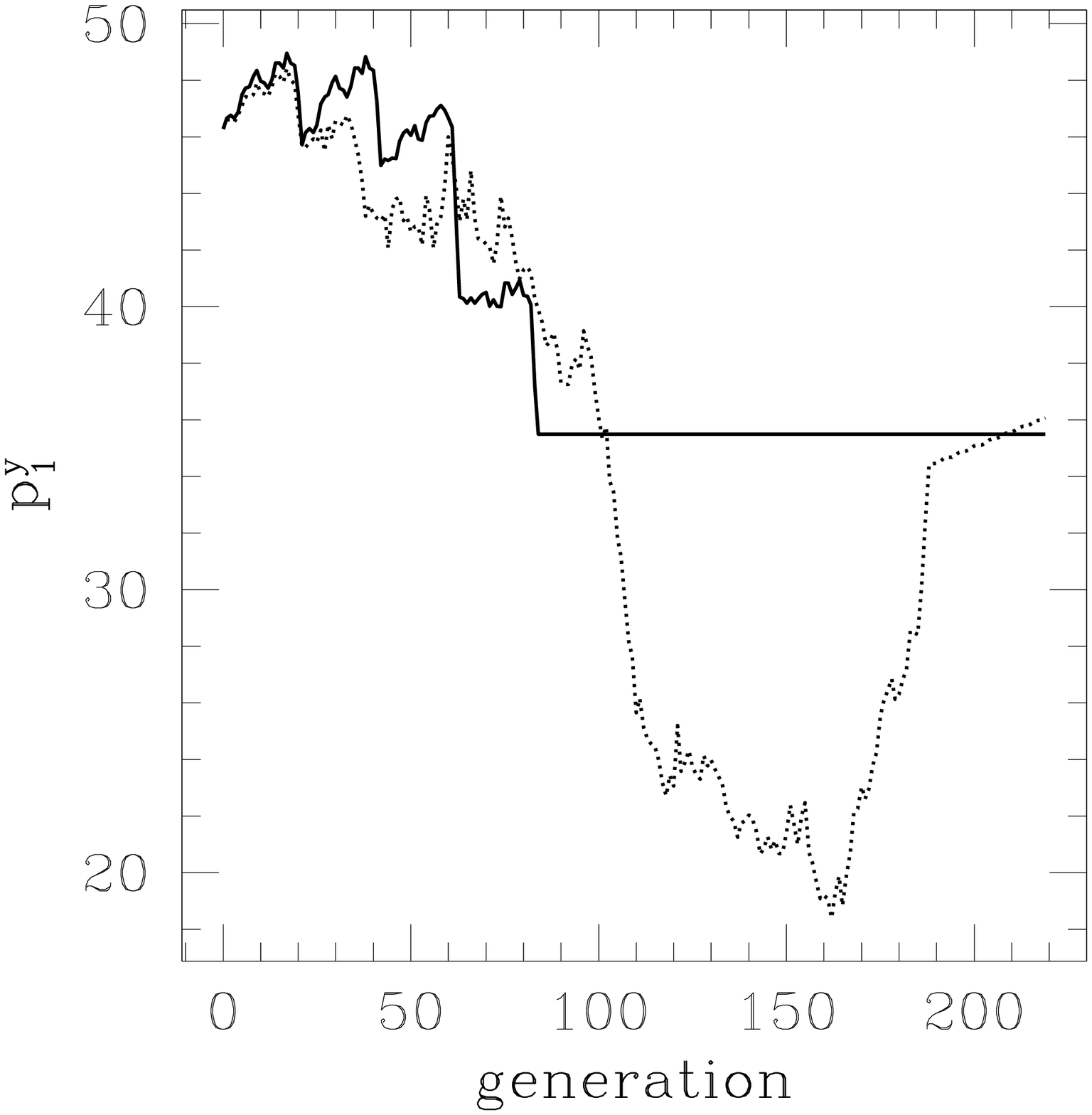,height=5cm}
\end{minipage}\hfill
\caption{The Minkowski functionals for the two ``Game of Life'' series
depicted  in  Figure~\ref{fig:gol}.  We  compare  the  length  of  the
circumference (left  panel) and  both components of  the corresponding
centroid                (middle                and               right
panel)\label{fig:golmink}.     ``thunderbirdfuse'':     solid    line,
``perturbed thunderbirdfuse'': dashed line.}
\end{figure}
\begin{figure}
\begin{minipage}[t]{.99\linewidth}
\centering
\epsfig{file=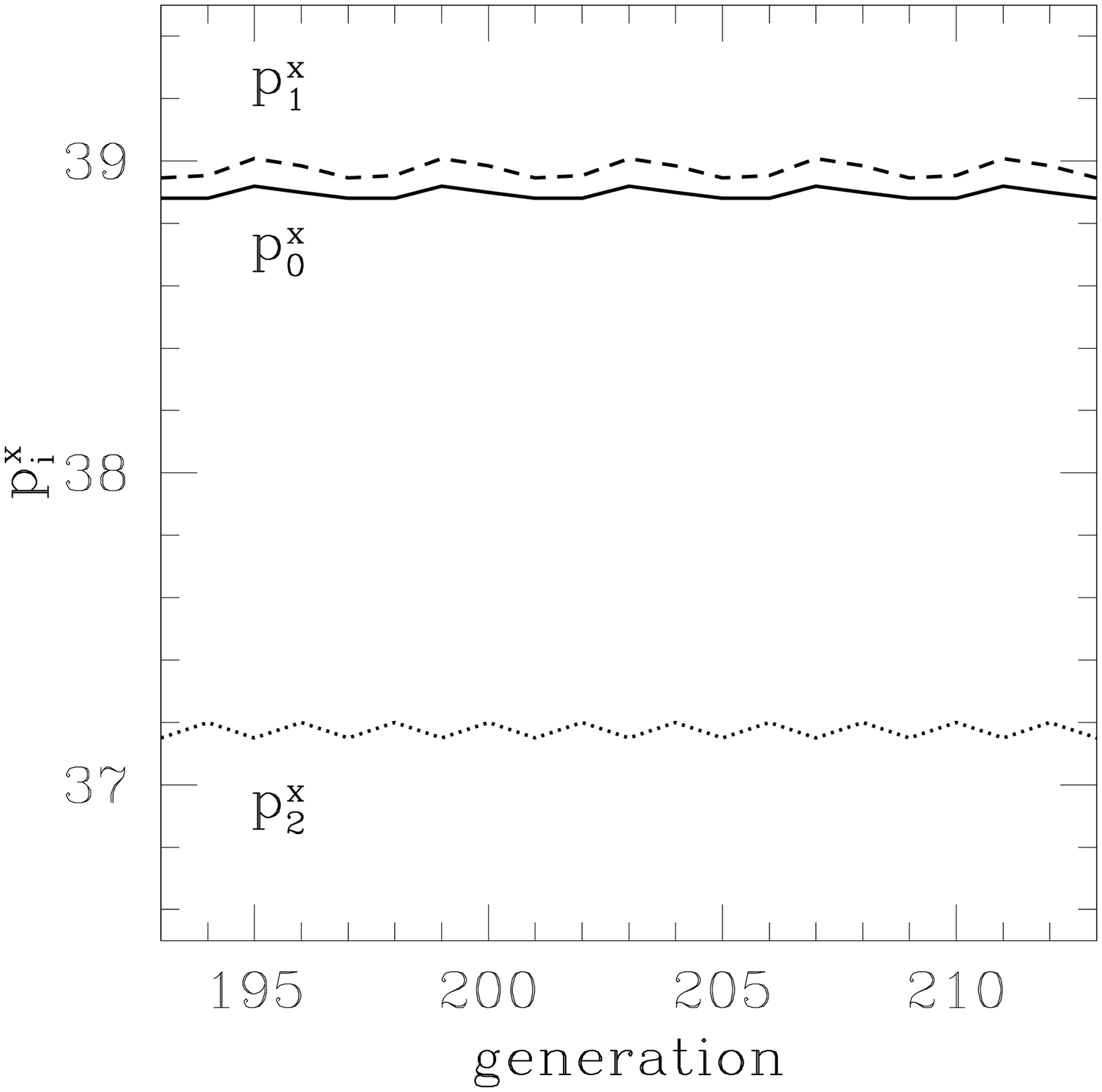,height=6cm}
\epsfig{file=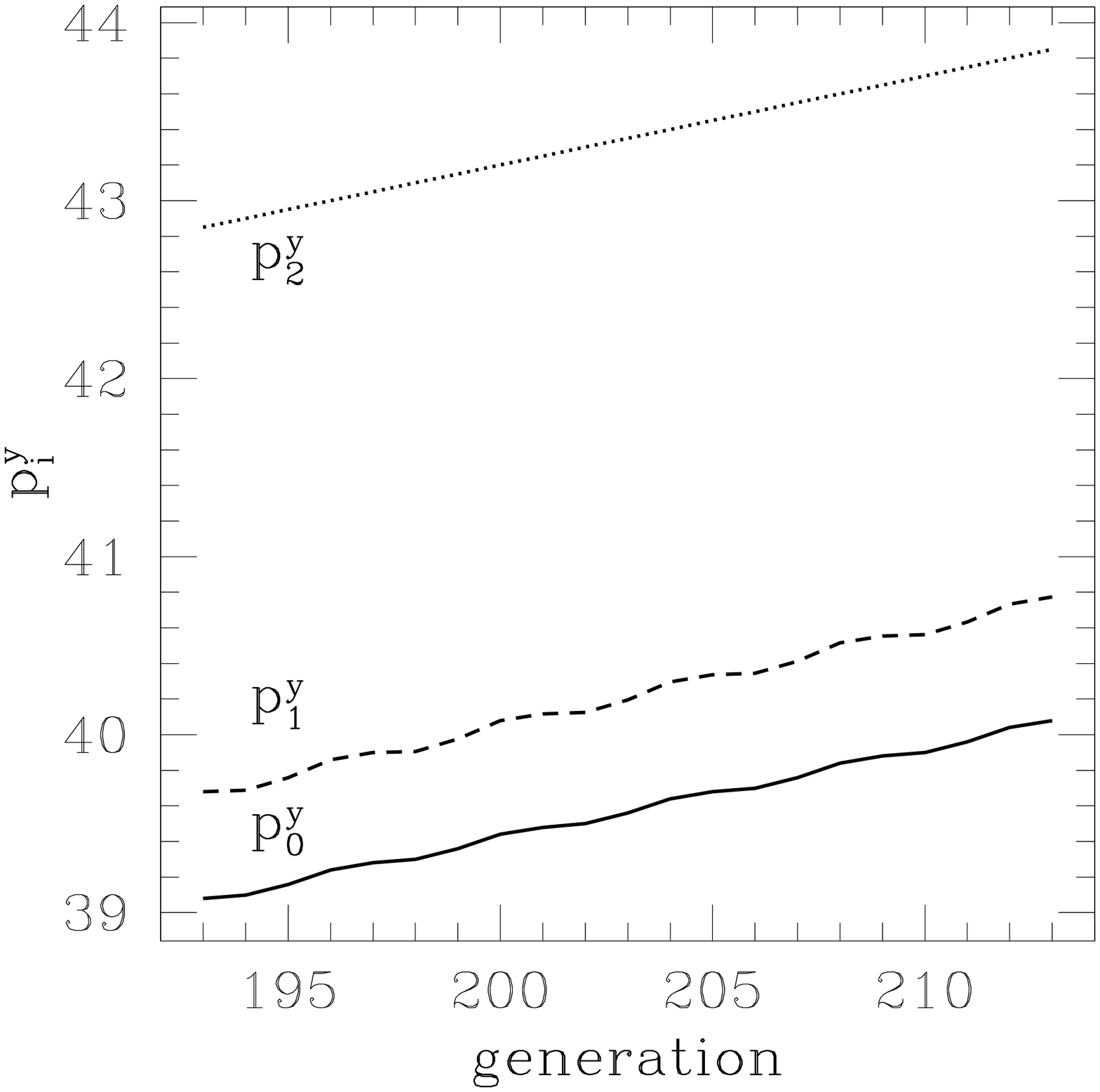,height=6cm}
\end{minipage}\hfill
\caption{The  interplay  of all  curvature  centroids  for the  second
``Game of Life'' series  depicted in Figure~\ref{fig:gol} allows us to
answer  questions  on  the  symmetry  and  movement  of  the  patterns
arising.\label{fig:golcc}}
\end{figure}
$C_1$  may   be  distinguished  from   $C_2$  by  the  value   of  the
isoperimetric ratio, $I(C_1)  =1$ (which holds only for  the disk) and
$I(C_2)>1$, where $I(K) \equiv  4 V_1^2(K)/[\pi V_0(K) V_2(K)] \geq 1$
for  convex  bodies.  The  centroids  of  $C_3$  and $C_4$  no  longer
coincide; in  $C_3$ they  fan out along  the single  mirror reflection
line, whereas  in $C_4$ they form  a triangle. The  clusters $C_5$ and
$C_6$ have the same three components, but in different configurations.
The scalar Minkowski  functionals of both $C_5$ and  $C_6$ differ from
the  ones  of  $C_1$--$C_4$  at  least  by  the  value  of  the  Euler
characteristic  $V_2$,  which  counts   here  the  number  of  cluster
components.  But the  scalar Minkowski  functionals are  unsuitable to
distinguish  $C_5$  from $C_6$  --  additivity  implies  $V_i (C_5)  =
V_i(C_6)$  --,  whereas the  centroids  discriminate  clearly.  \\  In
Figure  \ref{fig:gol}  we  illustrate  the order  parameters  using  a
dynamical   cellular   automaton    generated   by   the   ``Game   of
Life''~\cite{berlekamp:ways,gardner:life}.  We  consider two series of
patterns, the  ``thunderbirdfuse''~\footnote{available, e.g.  from the
library     of     ``Game     of     Life''     patterns     ``xlife''
(http://www.mindspring.com/\-$\sim$alanh/  life/).   }   and a  series
starting  with  a  slightly  perturbed initial  configuration  of  the
``thunderbirdfuse'',   with   one   point   added  (left   column   of
Figure~\ref{fig:gol}).   The ``thunderbirdfuse''  lights consecutively
the ``blinkers''  visible as bars.   The temporal sequence as  well as
the  spatial course  of the  ignitions is  reflected by  the Minkowski
functionals  (Figure~\ref{fig:golmink}) and  the  curvature centroids.
The  evolution  terminates with  ``traffic  lights'' oscillating  with
period  two   and  constant  values   of  the  scalar   and  vectorial
functionals.    \\   The   evolution   of   the   slightly   perturbed
\begin{table}
\centering
\setlength{\extrarowheight}{1pt}
\begin{tabular}{|l|l|l|}
\hline Minkowski functional & meaning & corresponding vector \\ \hline
 $V_0$                      & surface content & $\int_K \d^2 V \x$    \\ \hline
 $V_1$                      & one fourth of the length of perimeter & $\frac{1}{4}\int_{\partial K} \d^1 S \x$    \\ \hline
 $V_2$                      & Euler characteristic & $\frac{1}{2\pi}\int_{\partial K} \d^1 S\; c(\x) \x$    \\ \hline 
\end{tabular} 
\caption{The meaning  of the Minkowski functionals  in two dimensions,
$c(\x)$ is the local curvature. \label{tab:meaning}}
\end{table}
``thunderbirdfuse'' proceeds  rather differently.  The  early patterns
are  still comparable;  at  the  time, when  the  pattern reaches  the
modified pixel, the perturbed  ``thunderbirdfuse'' grows and attains a
constant  surface content from  generation 188  on.  However,  as seen
from  $\p_1$,  the pattern  keeps  changing  by  ejecting two  gliders
visible  in the  lower right  panel.  The  behaviour of  the curvature
centroids  is shown  in Figure~\ref{fig:golcc}:  The asymmetry  of the
pattern and its  temporal variation (the ``gliders'' move  one cell in
diagonal direction within four generations) is clearly recognizable.


\section{Applications to galaxy clusters}
\label{sec:app}
\begin{figure}
\begin{minipage}[t]{.33\linewidth}
\centering
\epsfig{file=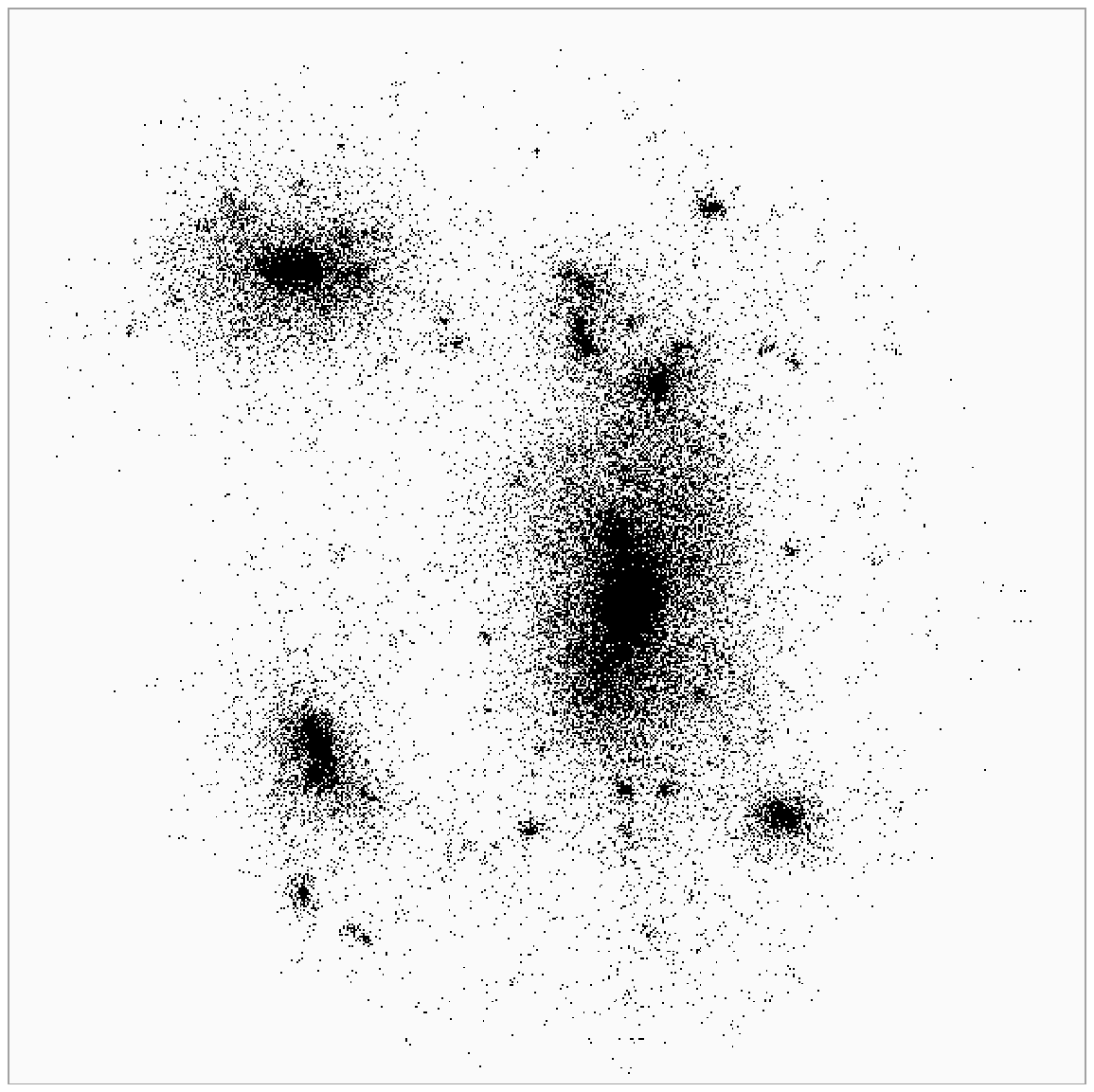,height=5cm}
\end{minipage}\hfill
\begin{minipage}[t]{.33\linewidth}
\centering
\epsfig{file=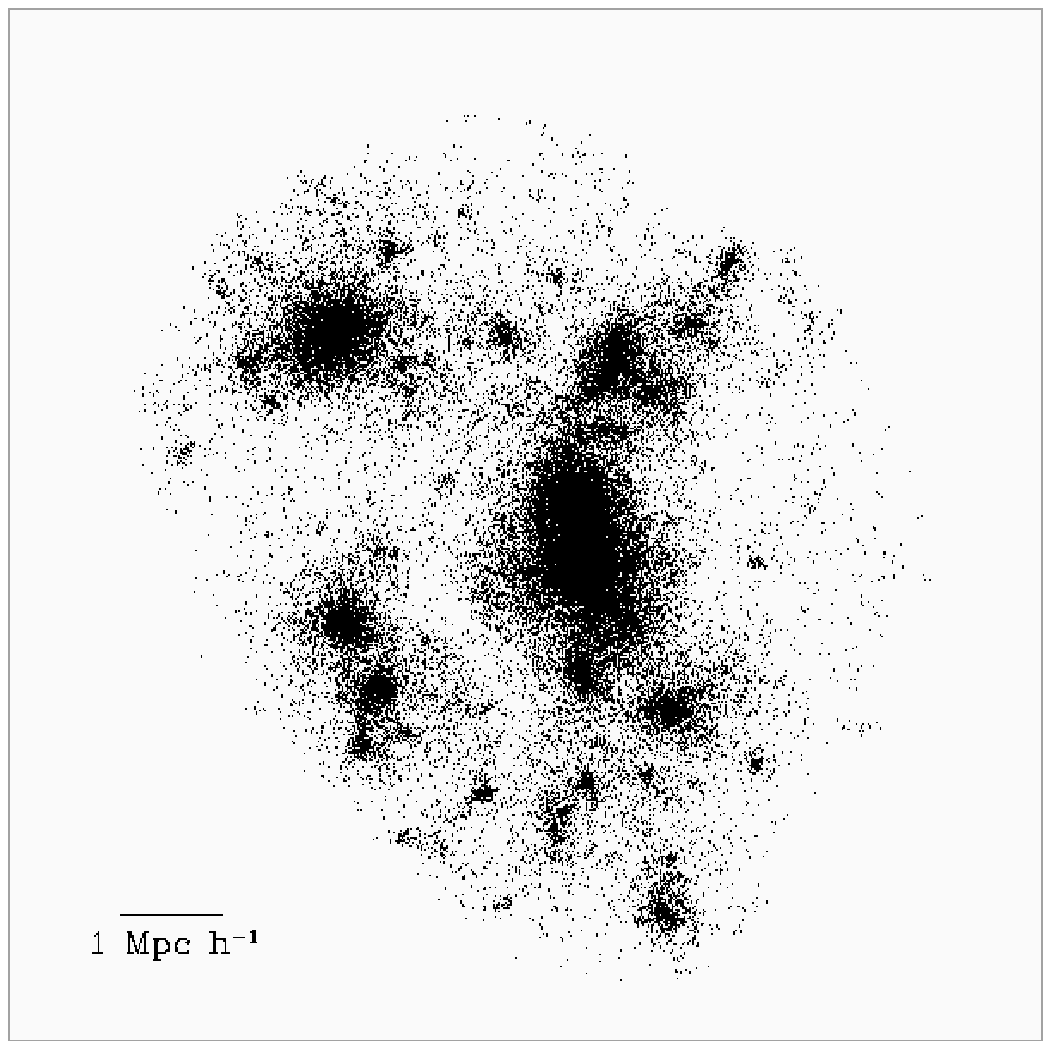,height=5cm}
\end{minipage}\hfill
\begin{minipage}[t]{.33\linewidth}
\centering
\epsfig{file=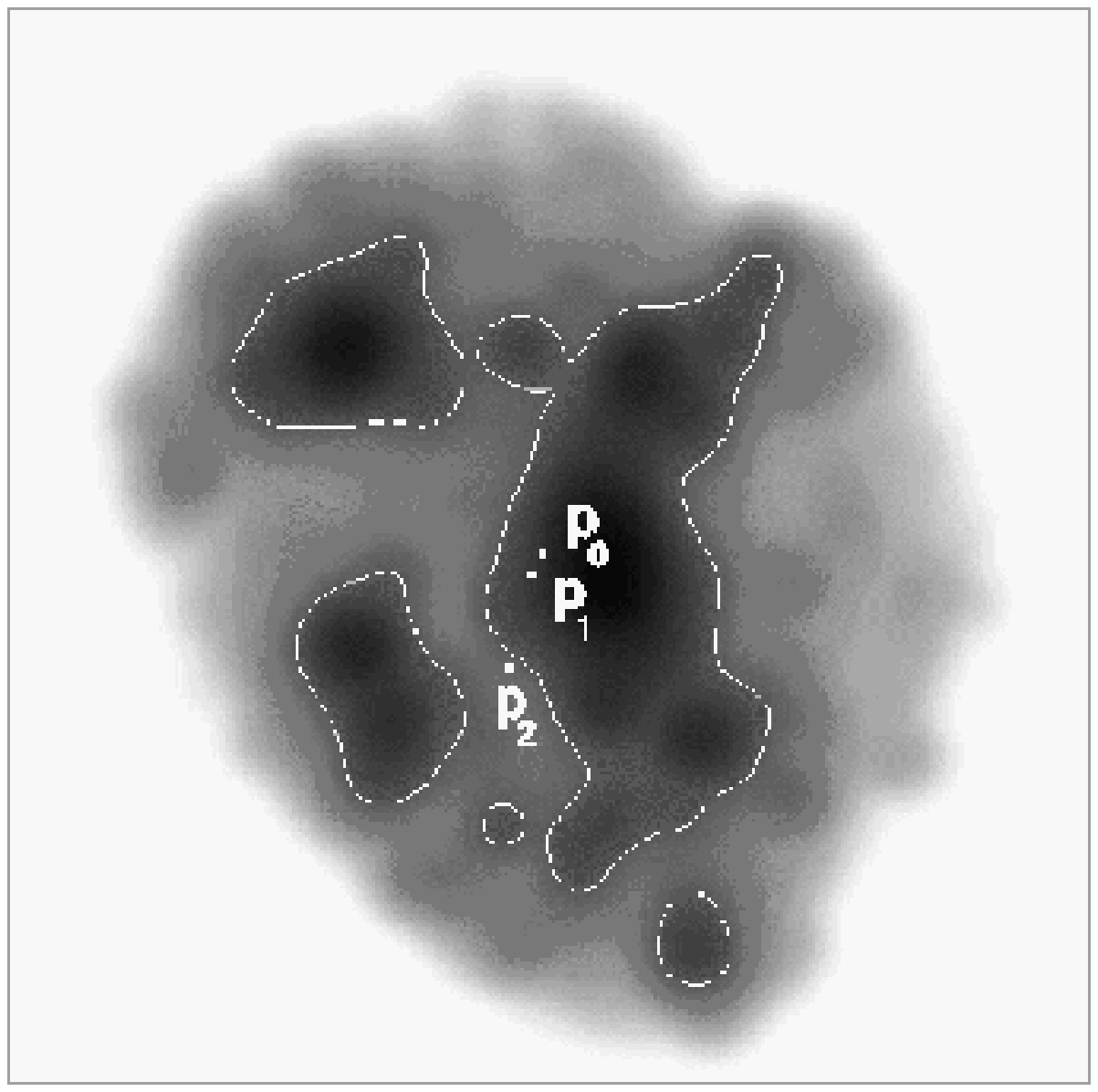,height=5cm}
\end{minipage}\hfill
\caption{ Example for the excursion  set approach.  The data points of
the cluster  $\tau$CDM (middle panel)  are smoothed using  a smoothing
length of $0.2\mpc$.  In the right  panel we show the contours and the
curvature centroids $\p_i$ of the excursion set at a density threshold
of $u  = 0.275  N/R_m^2$.  Here, $N$  denotes the number  of particles
within the cluster, $R_m$ quantifies the cluster scale via the maximum
distance     of     cluster     particles     from     the     cluster
center.\label{fig:constr} --  $h$ accounts for the  uncertainty in the
determination   of  the   Hubble   constant   $H$:  $   H   =  100   h
\frac{km/s}{Mpc}$.   --  The  left  panel  shows  the  counterpart  of
$\tau$CDM,  OCDM,   which  is  embedded  into   an  open  cosmological
model.}
\end{figure}
Evidently, observational  data of  galaxies within clusters  or X--ray
photon  maps  do not  provide  from the  outset  the  kind of  spatial
patterns required  for the application  of our morphometric  tools. To
start  with,  these  patterns  must  first  be  constructed  from  the
observational data; but  there is no canonical way  to proceed.  Here,
we employ two procedures, the  {\it excursion set method} and the {\it
Boolean grain method}  in order to further illustrate  our approach on
the basis of simulated and real cluster data.
\begin{figure}
\begin{minipage}[t]{.99\linewidth}
\centering
\epsfig{file=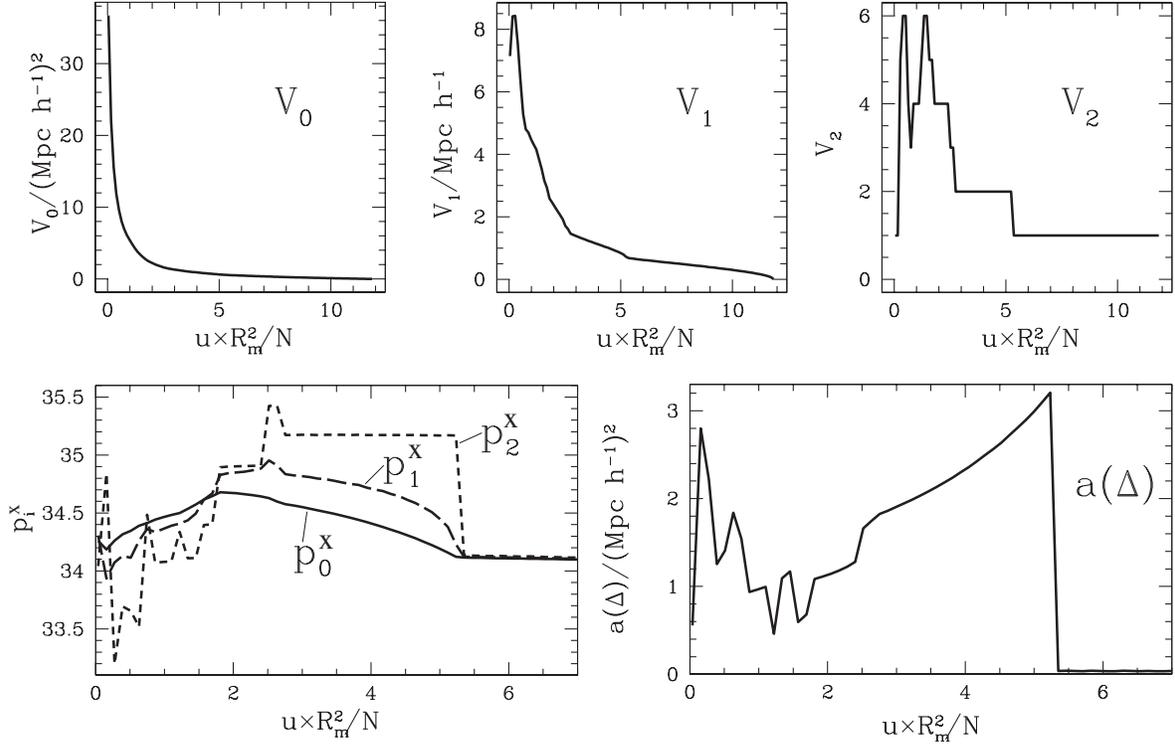,height=10cm}
\end{minipage}\hfill
\caption{
  First row: the scalar  Minkowski functionals of the cluster depicted
  in Figure \ref{fig:constr} vs. the density threshold $u$ ($u$ is
  given in  units of the mean  density within the cluster,  $N$ is the
  number of  cluster particles, $R_m$ the maximum  distance of cluster
  particles from the center of  mass of all cluster particles). Second
  row: one component  of the curvature centroids (left  panel) and the
  area  of the  triangle  spanned by  the  curvature centroids  (right
  panel).\label{fig:bsp}}
\end{figure}

\subsection{The excursion set method}

We smooth  the projected galaxy positions  or the pixels  of the X-ray
photon maps  with a Gaussian kernel.  The  smoothing length determines
the  scale of  interest.  Then  we  construct the  excursion sets  and
investigate their topology and  geometry using the Querma{\ss} vectors
and   the  Minkowski   functionals~\cite{schmalzing:beyond}.    \\  We
illustrate this procedure by comparing two simulated clusters (part of
the              GIF--simulations,              {\em              cf.}
\cite{bartelmann:arcIV,beisbart:app});  they   start  with  comparable
initial conditions and evolve within different Friedmann--Lema\^\i tre
models as  cosmological background,  thus exemplifying the  imprint of
the background.  We demonstrate  our method in Figure \ref{fig:constr}
using   the   cluster    $\tau$CDM   from   an   Einstein--de   Sitter
model~\footnote{A $\tau$CDM model  is a variety of a  Cold Dark Matter
structure  formation  scenario embedded  into  an Einstein--de  Sitter
background.}  and a smoothing length  of $0.2\mpc$. The results of the
Minkowski analysis are depicted  in Figure \ref{fig:bsp}.  We plot the
scalar Minkowski  functionals vs.  the density  threshold defining the
excursion sets: an averaged and smoothed density profile is encoded in
the first  Minkowski functional $V_0$;  a comparison of the  square of
$V_1$ and  $V_0$ quantifies how  crooked the isodensity  contours are.
The Euler characteristic $V_2$ counts the components (minus the number
of  eventual  holes)  of  this  substructure--rich  cluster.   \\  The
components $p_i^x$ of the curvature  centroids are shown in the fourth
panel of  Figure~\ref{fig:bsp}.  They wander  in space if  the density
threshold is  varied.  This indicates that  morphological features are
shifting.  Another way  to make the centroids more  illustrative is to
consider  the triangle  spanned by  the centroids  and to  compute its
volume  and  the  length  of  its perimeter  (fifth  panel  of  Figure
\ref{fig:bsp}).  These quantities tell us how symmetric the isodensity
contours are.\\ To compare this  cluster to OCDM, its counterpart in a
low--density   background   model,   we  condense   the   morphometric
information  present in  the Minkowski  functionals and  the centroids
into  a few dimensionless  descriptors, by  taking into  account, that
both  clusters have  different scales  and numbers  of  particles.  We
define for a function $f$ an average over density thresholds via:
\begin{equation}
\daverage{f} = \frac{1}{u_{max}-u_{min}}\int_{u_{min}}^{u_{max}} \d u f(u)\;\;\;
\end{equation}
\begin{table}
\centering
\setlength{\extrarowheight}{1pt}
\begin{tabular}{|l|l|l|l|l|}
\hline
$\sigma/\mpc$ & parameter& definition & $\tau$CDM  & OCDM  \\\hline\hline
\multirow{4}{10mm}{0.2}   &  c               &  $\sqrt{\daverage{(\chi-1)^2}}$  &   $3.20$        &    $2.54$      \\ \cline{2-5}
                           & $s_0/R_{cl}^2$   &  $\daverage{4V_0(\triangle(\p))}/R_{cl}^2$      &   $8.78\times 10^{-4}$    &    $6.31\times 10^{-4}$        \\ \cline{2-5}
                           & ${\rm shift}_0/R_{cl}$ &         $\int_{u_{min}}^{u_{max}} \left| \frac{\d  \p_i}{\d u} \right|\d u /R_{cl}$       \;\;\;.&  $3.85\times 10^{-1}$   &    $4.54\times 10^{-1}$        \\ \hline\hline
\end{tabular} 
\caption{  The  definitions of  the  condensed  descriptors and  their
values  for  the  clusters  $\tau$CDM  and  OCDM  depicted  in  Figure
 6. $4V_0(\triangle(\p))$  is the circumference  of the
triangle          spanned           by          the          curvature
centroids\label{tab:tauopen}.  $R_{cl}$ is the  scale of  the cluster,
determined via the area of the  convex hull of the cluster $A_h$: $\pi
R_{cl}^2= A_h$. $\sigma$ denotes the smoothing length.}
\end{table}
and consider  the {\it clumpiness},  the {\it symmetry  parameter} and
the {\it  shift of morphological  properties} as defined in  the third
column   of   Table~\ref{tab:tauopen}~\footnote{Here,  $u_{min}$   and
$u_{max}$ denote the maximum  density found within cluster and outside
the cluster, respectively, where ``outside the cluster'' means farther
away  from  the  cluster   center  than  any  cluster  point}.   These
parameters display the imprint of subclumps, the symmetry and shift of
the curvature centroid $\p_2$ and are the larger the more substructure
the  cluster exhibits.  As  visible from  Table~\ref{tab:tauopen}, the
cluster $\tau$CDM  owns more  substructure with respect  to clumpiness
and  symmetry.   A statistical  comparison  of  these descriptors  for
larger  numbers of  clusters simulated  in different  models  yields a
typical  cluster morphology  and  may help  to  constrain the  present
values of the cosmological parameters~\cite{beisbart:app}.
\subsection{The Boolean grain method}
\begin{figure}
\begin{minipage}[t]{.49\linewidth}
\centering
\epsfig{file=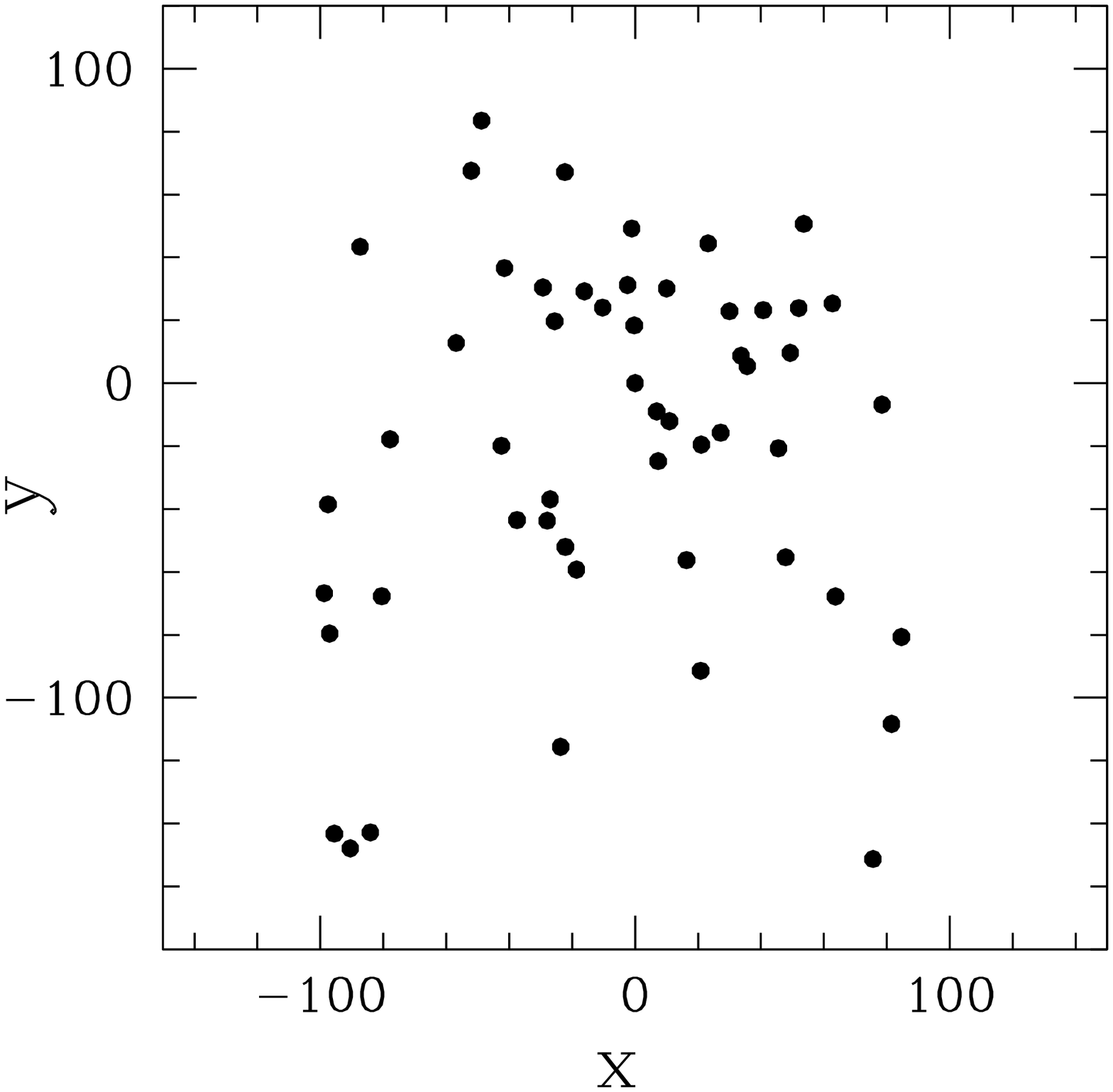,height=7cm}
\end{minipage}\hfill
\begin{minipage}[t]{.49\linewidth}
\centering
\epsfig{file=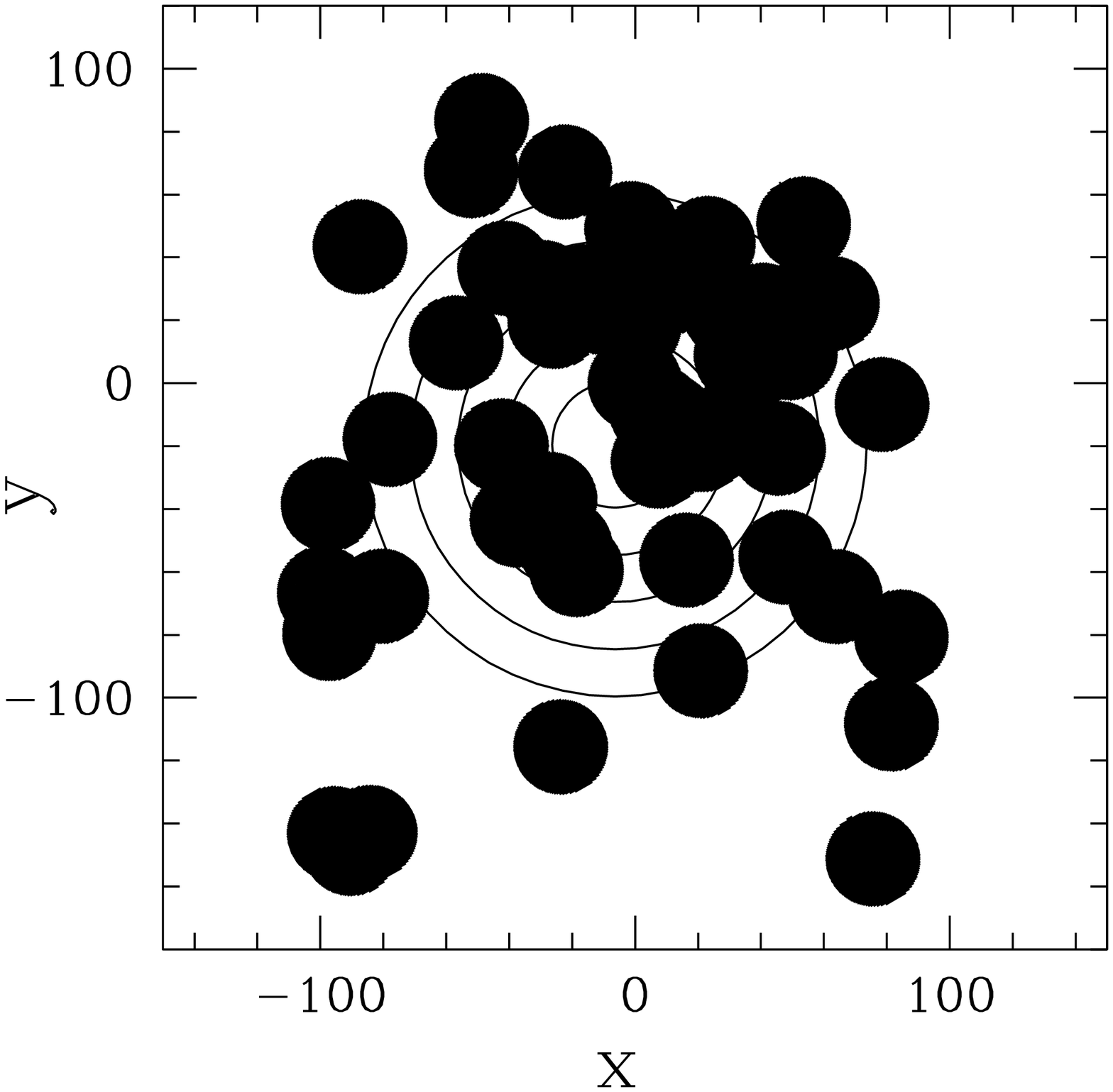,height=7cm}
\end{minipage}\hfill
\caption{
  The  cluster  population  of   Cl  0016+161  (left  panel)  and  the
  associated Boolean grain model  with circular windows (right panel). 
  The radius of the Boolean grains is $15$ arcseconds.
  The units are arcseconds.\label{fig:method}}
\end{figure}
\begin{figure}
\begin{minipage}[t]{.33\linewidth}
\centering
\epsfig{file=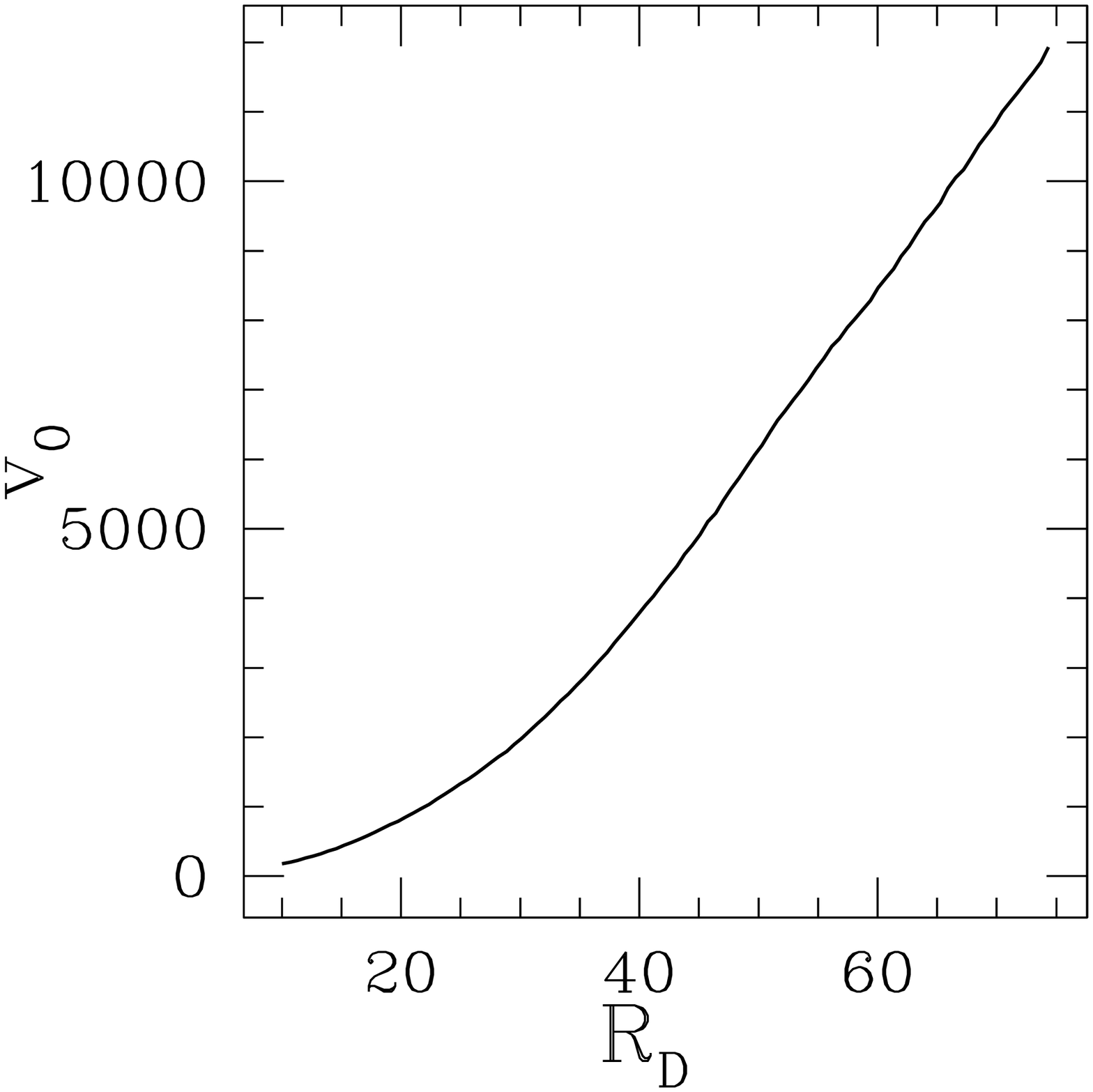,height=5cm}
\end{minipage}\hfill
\begin{minipage}[t]{.33\linewidth}
\centering
\epsfig{file=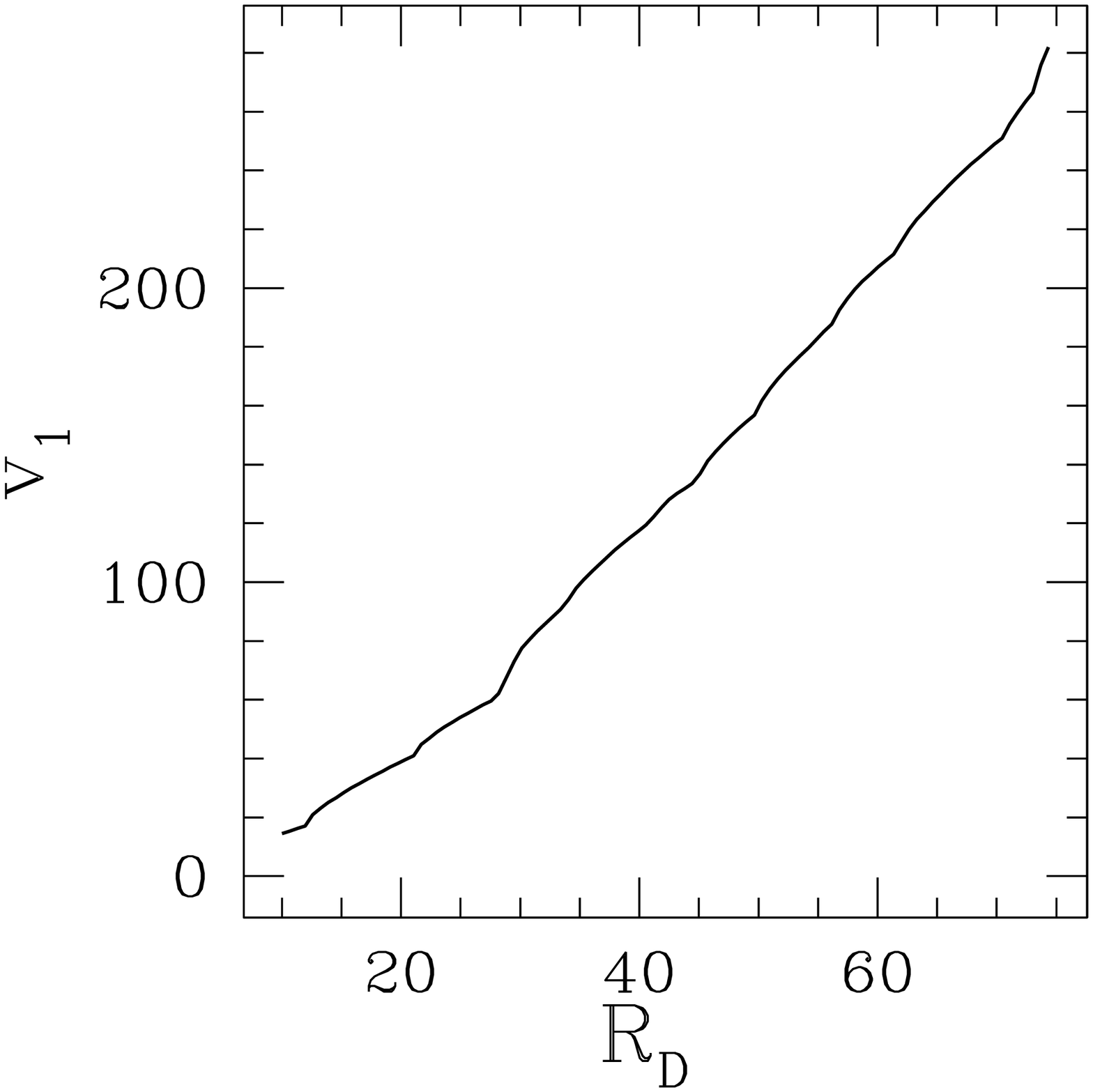,height=5cm}
\end{minipage}\hfill
\begin{minipage}[t]{.33\linewidth}
\centering
\epsfig{file=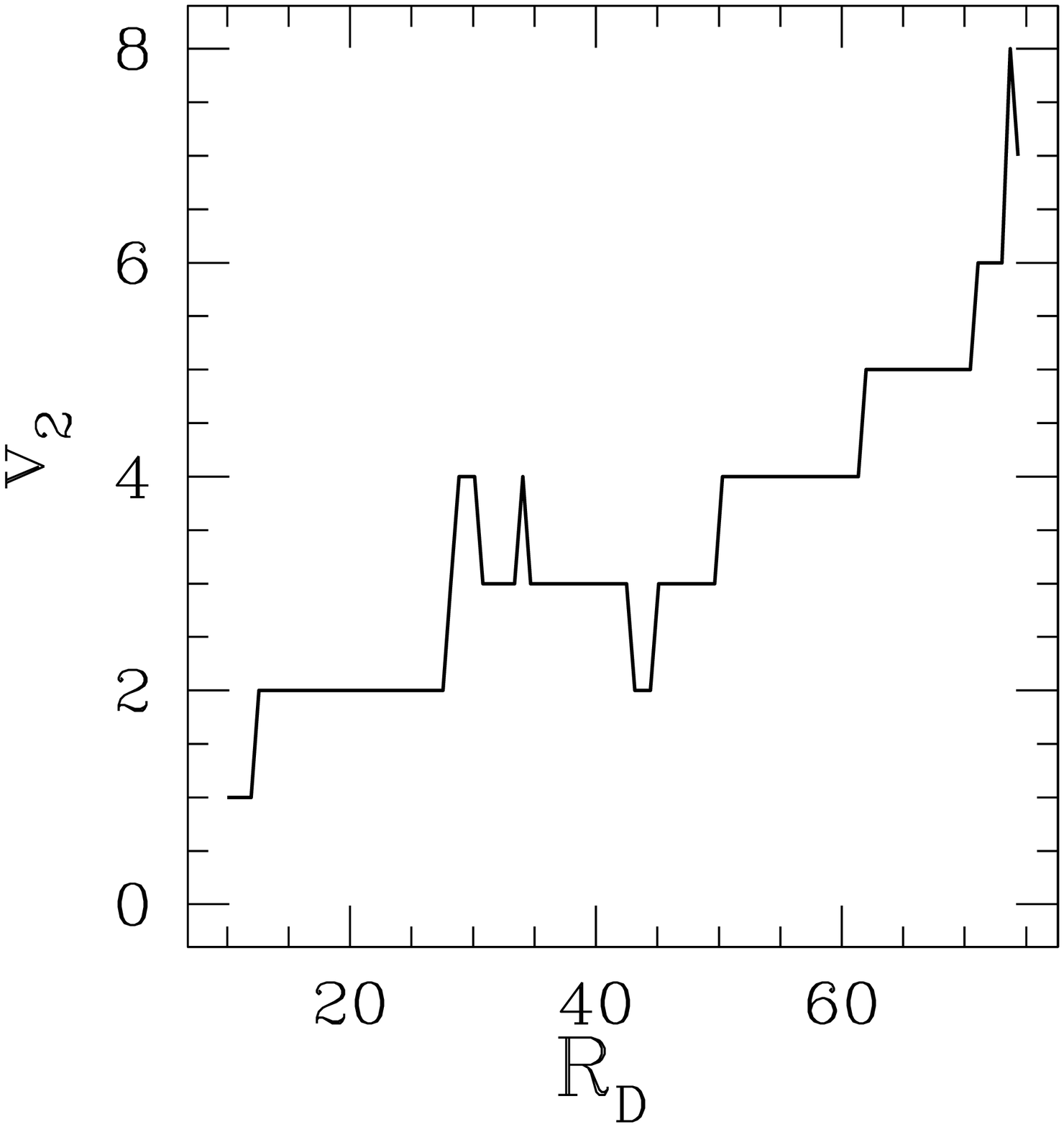,height=5cm}
\end{minipage}\hfill
\caption{
  The scalar Minkowski functionals for the cluster Cl 0016+161 vs.
  the radius $R_D$ (in arcseconds)  of the sampling window. The radius
  of the grains is fixed ($15''$), the Minkowski functionals $V_i$ are
  given in units of arcsec$^{2-i}$.\label{fig:rc_box}}
\end{figure}
The Boolean  grain method decorates each  point $\x_i$ of  a point set
with a ball $B_{\x_i}(r)$ of radius  $r$. The union set of these balls
is diagnosed  by computing the Minkowski functionals  and centroids as
functions  of the  radius. To  probe the  morphology of  an individual
cluster locally,  we place a window  $\CD$ on the  cluster center (the
center  of  the point  set)  and  study  the centroids  of  $\bigcup_i
B_{\x_i}(r)\cap \CD$. Inflation of the sampling window $\CD$ allows us
to explore different regions of the cluster. We illustrate this method
in  Figure  \ref{fig:method}. Note,  that  also  galaxies outside  the
window   contribute~\footnote{Computational   details  are   described
in~\cite{beisbart:quer}.}.   \\  To  provide  a concrete  example,  we
investigate  the  cluster Cl  0016+161  at  $z\sim  0.54$ observed  by
Belloni   and   R{\"o}ser~\cite{belloni:cluster}   with  the   angular
positions  and   spectral  properties  of  galaxies  in   a  field  of
$3.5'\times 5'$  and brighter than  $R=23.5 {\rm mag}$.   Galaxies are
\begin{figure}
\begin{minipage}[t]{.99\linewidth}
\centering
\epsfig{file=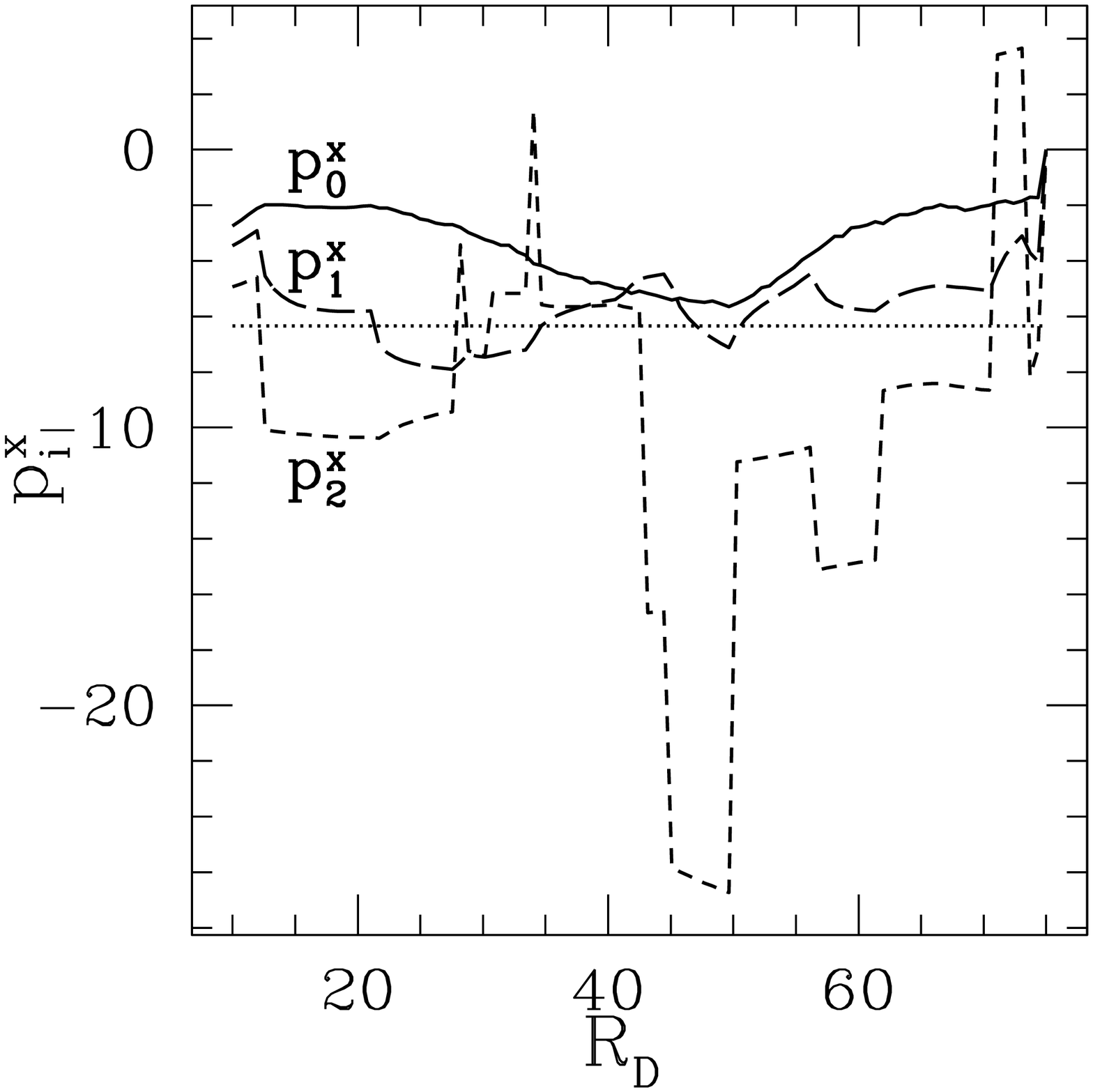,height=7cm}
\end{minipage}\hfill
\caption{
  The  behaviour of the  curvature centroids  ($p_i^x$--component) for
  the  Boolean  grain model  in  Figure  \ref{fig:method} vs.  the
  radius of  the window  (in units of  arcseconds). The  straight line
  indicates the cluster center.\label{fig:rc_box_p}}
\end{figure}
considered  as  cluster  members,  if  their redshifts  lie  within  a
predetermined       range\footnote{More      precisely,      following
\cite{belloni:cluster} elliptical  and E+A galaxies  are considered as
cluster  members, if  for  their redshift  $z$: $0.525<z<0.575$.   For
spiral and  irregular galaxies the redshift  range is $0.515<z<0.585$.
We  consider only  galaxies  whose morphological  type was  determined
(Table 5 in \cite{belloni:cluster}).}.  We are left with $53$ galaxies
in  this  field,  as  shown  in Figure  \ref{fig:method}.\\  We  place
spherical windows $\CD= B_{R_D}(\cv)$  of different sizes $R_D$ on the
center of  mass $\cv$  of the  point set.  The  radius of  the Boolean
grains is  held fixed at  $15''$.  In Figure \ref{fig:rc_box}  we show
the behaviour  of the scalar  Minkowski functionals.  The area  of the
union set of  balls inside the window $\CD$  increases almost linearly
with $R_D$.  A homogenous  galaxy distribution would yield a parabolic
dependence, but here  the galaxy density drops at  larger scales.  The
small peaks in the Euler characteristic typically arise when two parts
of  a  connected subcluster  enter  the  window simultaneously.\\  The
curvature  centroids are more  sensitive to  geometrical anisotropies.
The components $p_i^x$ are depicted in Figure \ref{fig:rc_box_p}.  Two
features  are clearly visible:  the centroids  vary when  the sampling
window  grows  and,  at  a  fixed scale  of  the  window,  significant
differences between  the curvature centroids arise.   The variation of
the  centroids is  clearly  correlated with  subclusters entering  the
window,  e.g.   the wandering  of  $\p_1$  and  $\p_0$ towards  higher
$x$--values   reflects  the  large   cluster  component   in  positive
$x$--direction of  the center.  \\  To analyze the morphology  of this
cluster more in detail, we  simulate a reference model for comparison.
The simplest way to generate  a reference cluster is an inhomogeneous,
but              spherically             symmetric             Poisson
process~\cite{daley:point-processes}.   We   determine  the  projected
density profile  of the cluster non--parametrically  around the center
of mass  using a  binning, and simulate  $100$ Poisson  clusters using
this profile (the method  for simulating inhomogeneous point processes
is  described e.g.  in  ~\cite{stoyan:fractals}).  The  mean Minkowski
functionals and the centroids for the model clusters with their r.m.s.
fluctuations as well as the  results for the real cluster are depicted
in  Figure \ref{fig:poissonvgl}.\\  The  scalar Minkowski  functionals
deviate only slightly from the  Poisson model.  On the other hand, the
curvature  centroids  reveal  significant  deviations (second  row  of
Figure~\ref{fig:poissonvgl});  in  particular,  the $y$--component  of
$\p_0$ (right panel) is shifted away  from the true center of mass for
relatively small values of the window scale, reflecting the fact, that
there are more galaxies above than beneath the cluster center.

\begin{figure}
\begin{minipage}[t]{.33\linewidth}
\centering
\epsfig{file=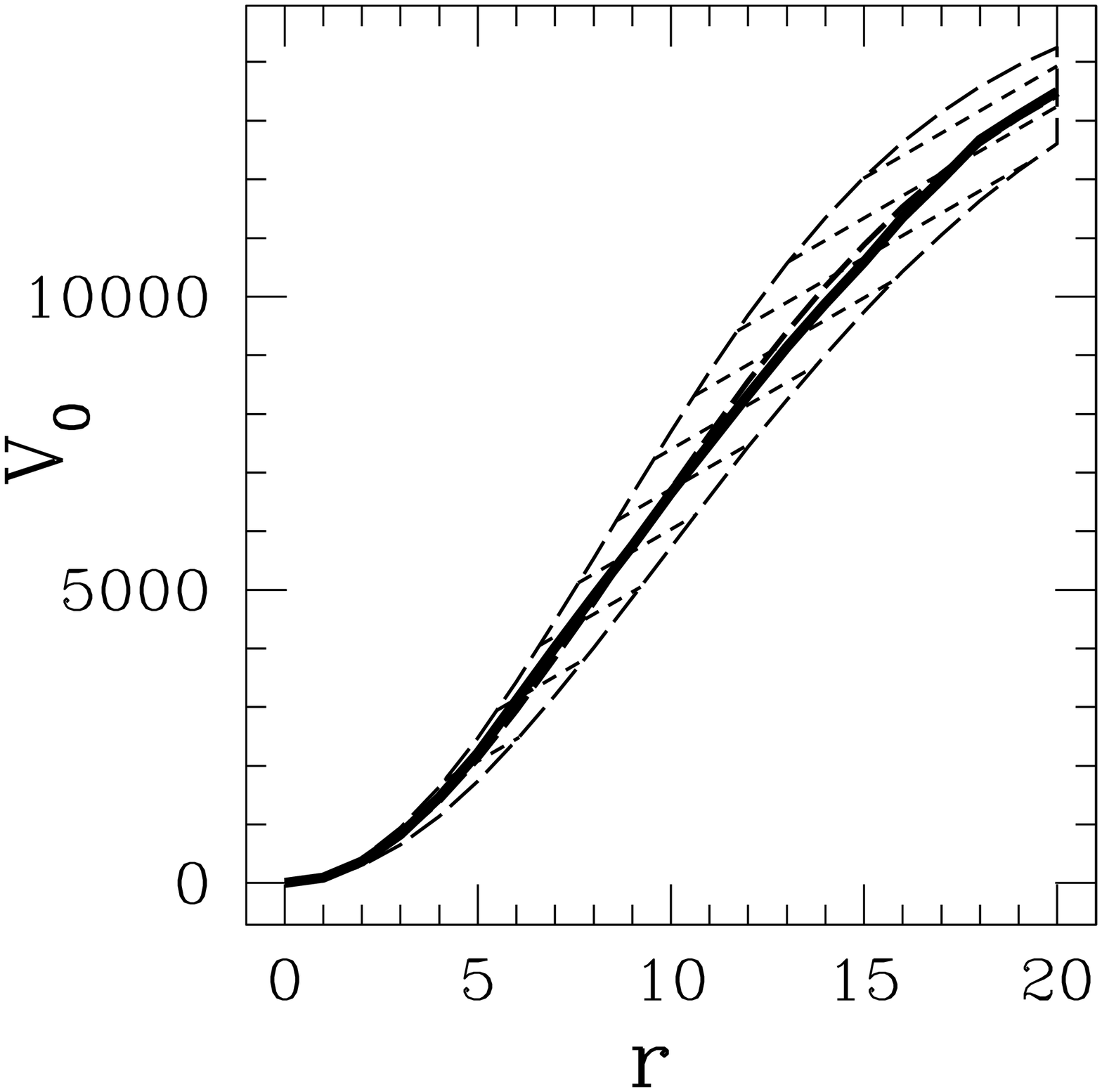,height=5cm}
\end{minipage}\hfill
\begin{minipage}[t]{.33\linewidth}
\centering
\epsfig{file=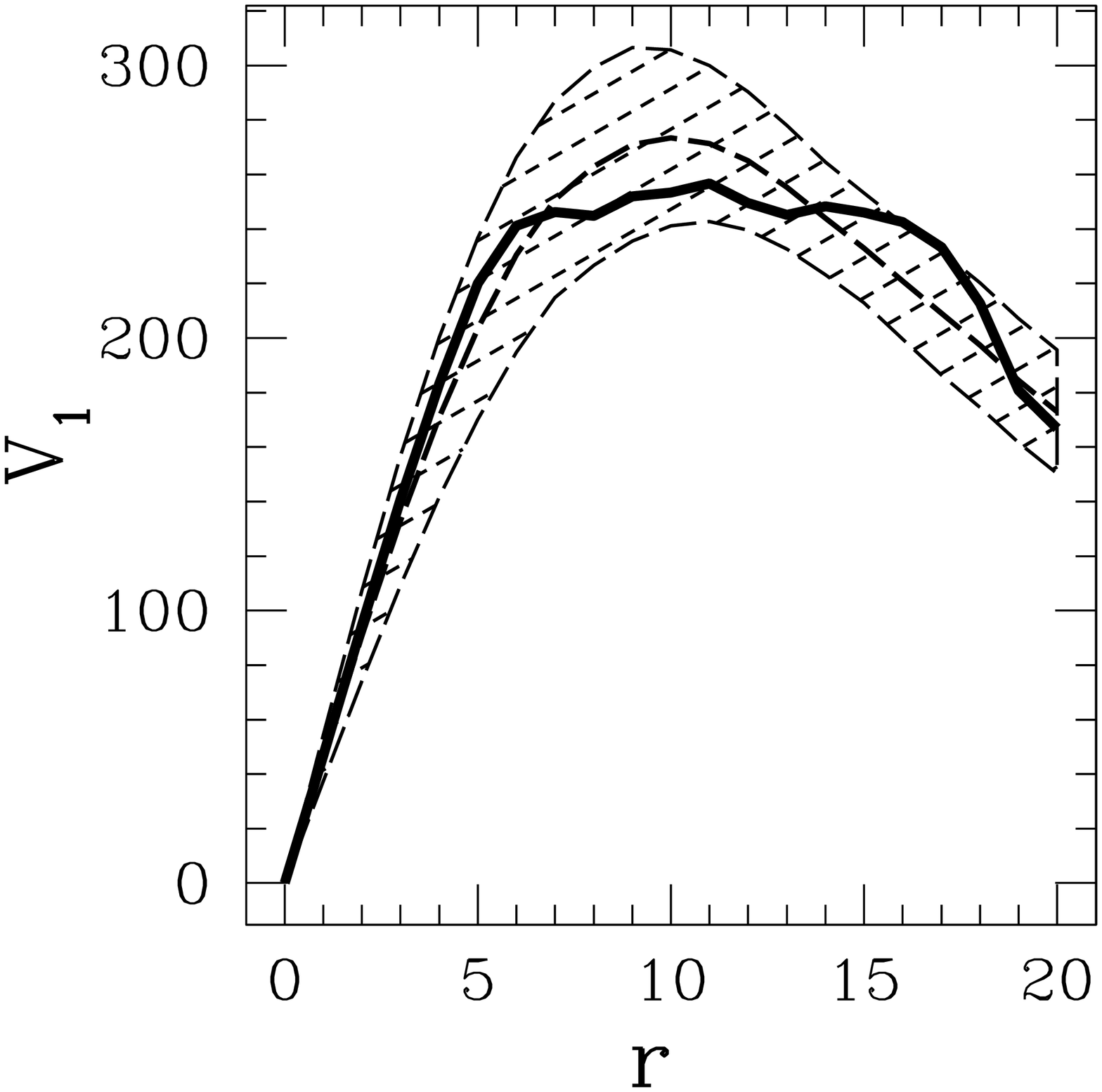,height=5cm}
\end{minipage}\hfill
\begin{minipage}[t]{.33\linewidth}
\centering
\epsfig{file=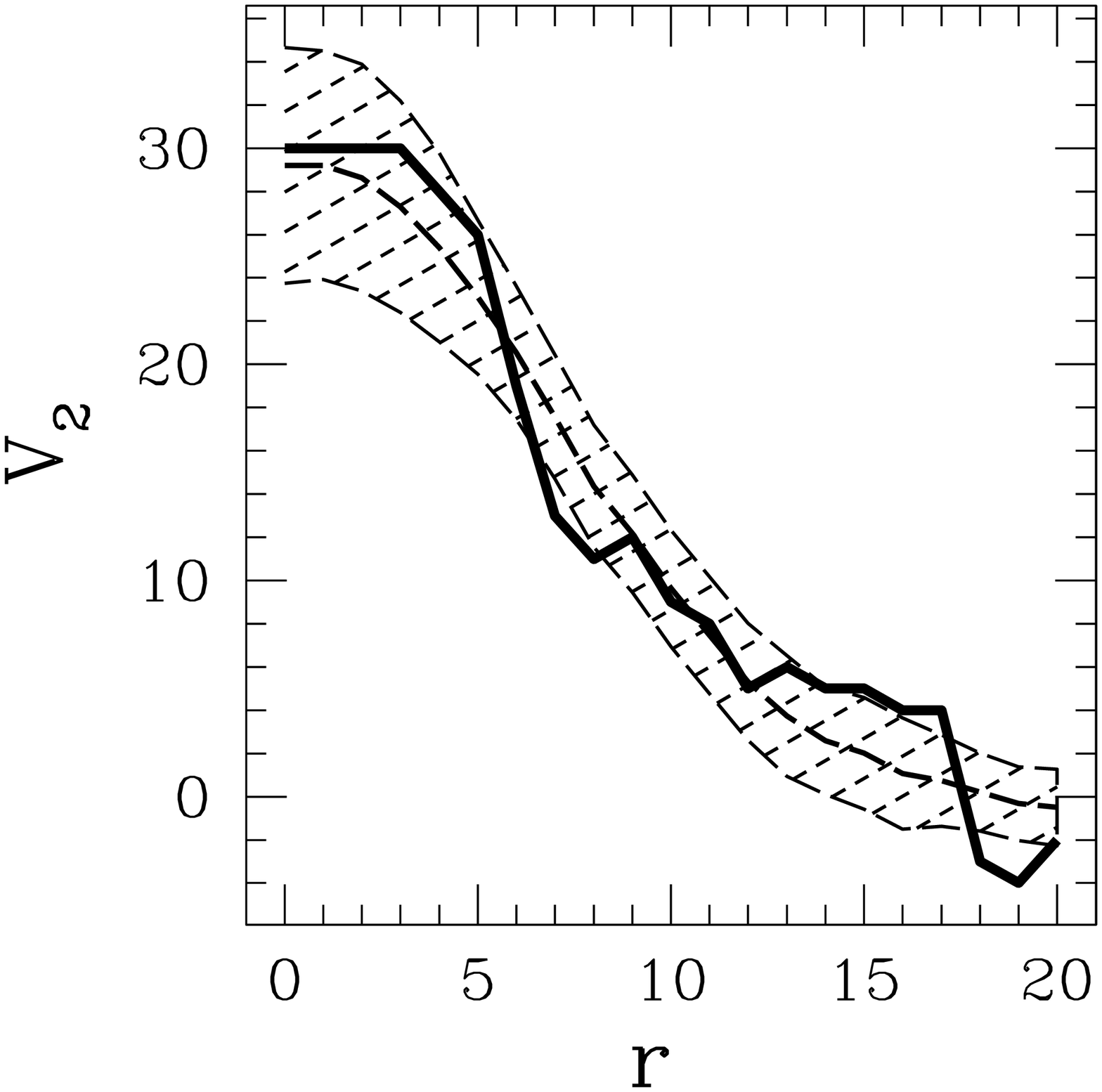,height=5cm}
\end{minipage}\hfill
\begin{minipage}[t]{.33\linewidth}
\centering
\epsfig{file=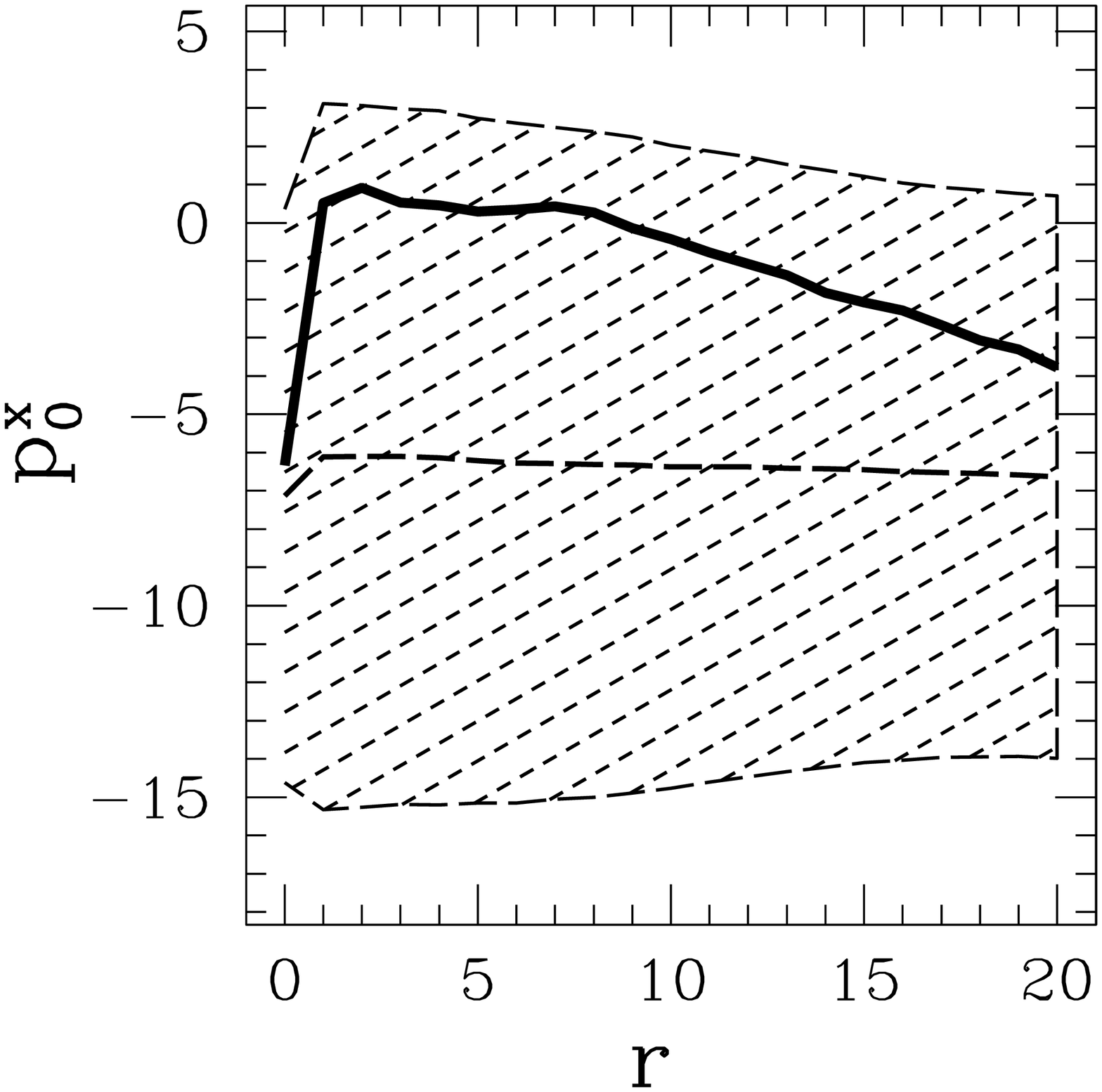,height=5cm}
\end{minipage}\hfill
\begin{minipage}[t]{.33\linewidth}
\centering
\epsfig{file=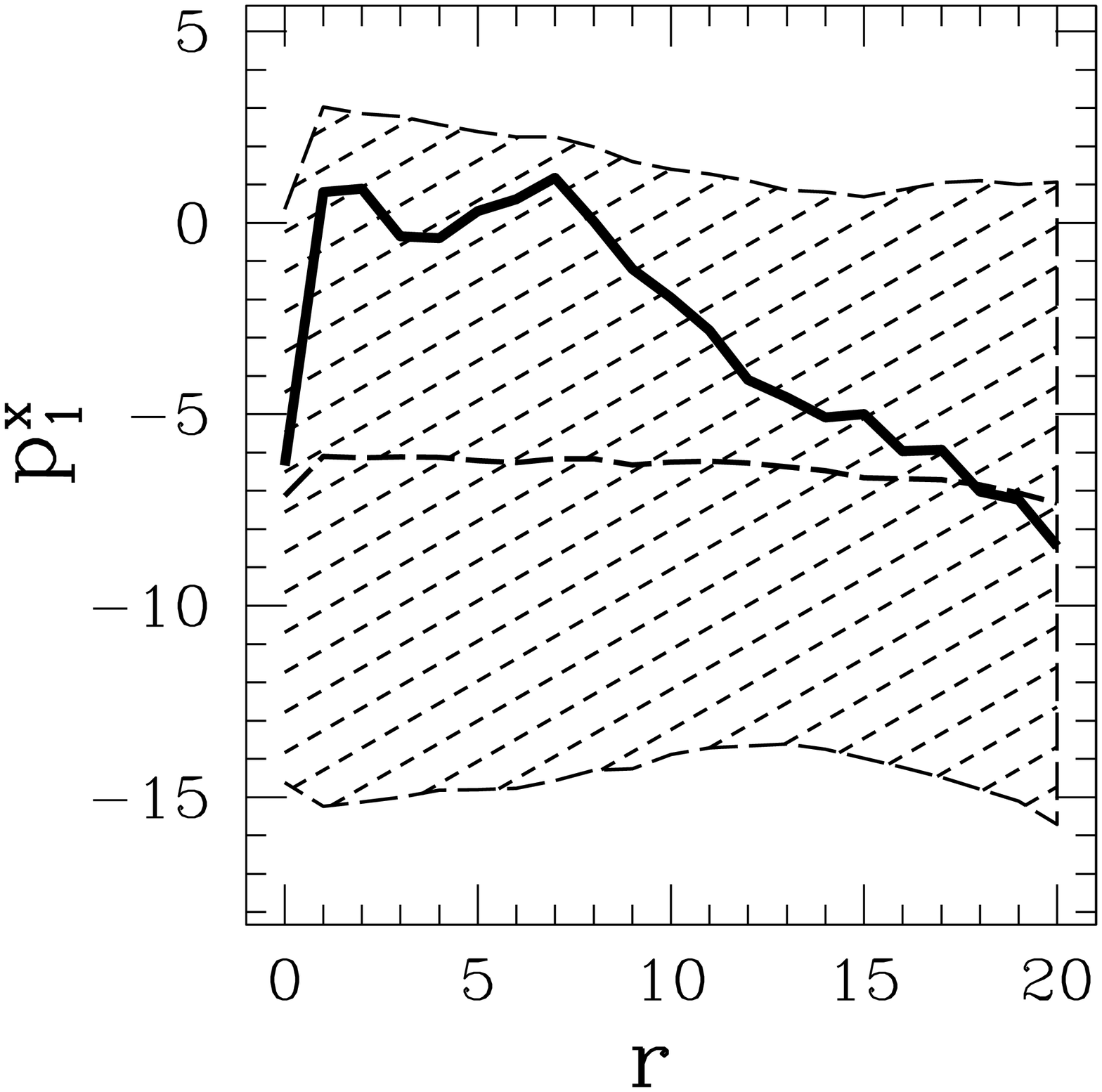,height=5cm}
\end{minipage}\hfill
\begin{minipage}[t]{.33\linewidth}
\centering
\epsfig{file=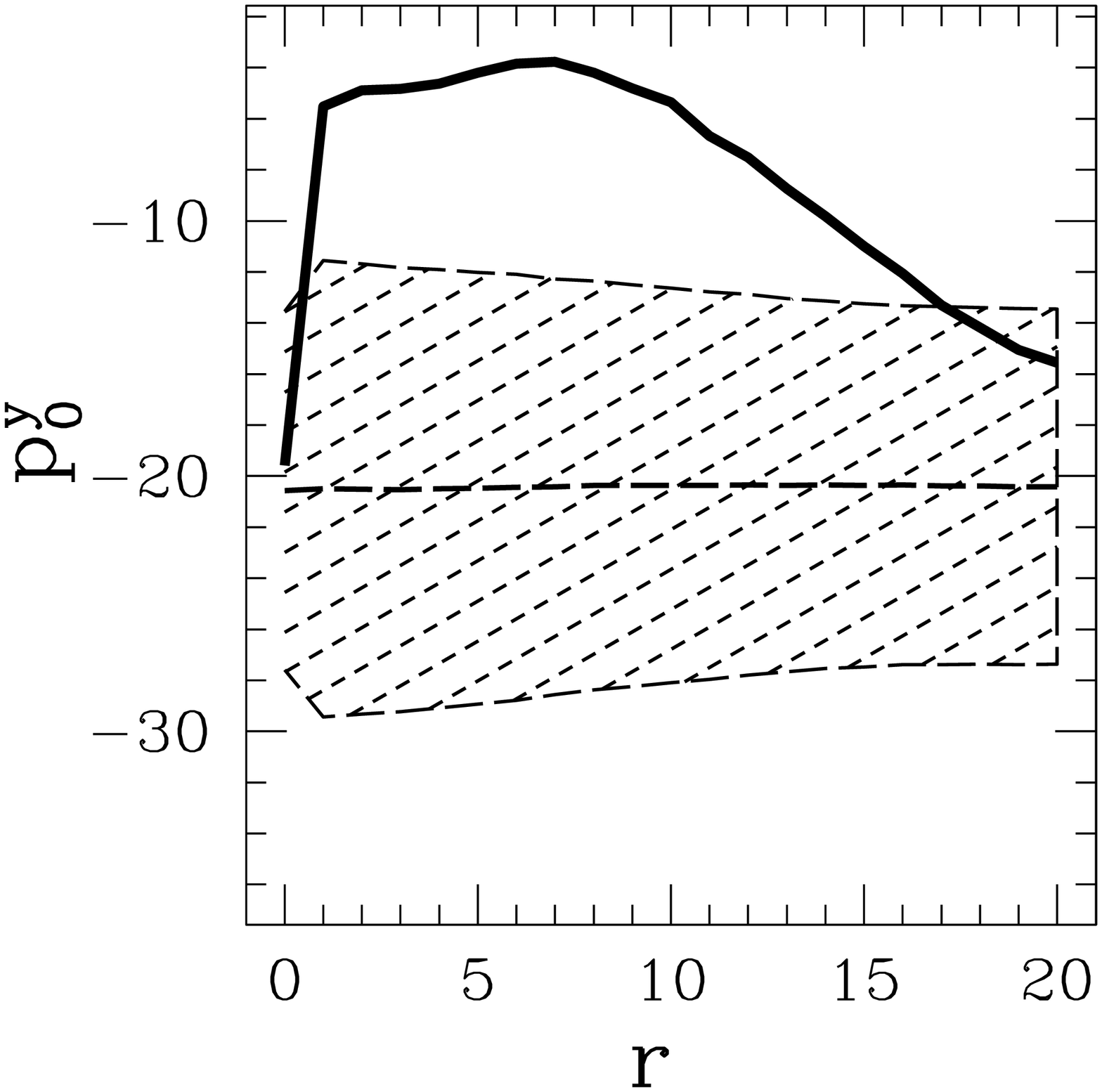,height=5cm}
\end{minipage}\hfill
\caption{  The Minkowski  functionals  (first row)  and components  of
different curvature centroids (second row) for the cluster Cl 0016+161
(solid line) and for a set of Poissonian random clusters following the
same density profile  as Cl 0016+161 (dashed line)  vs.  the radius of
the Boolean grains  in units of arcseconds.  The  radius of the window
is    fixed   at    $69''$.    The    shaded   areas    indicate   the
one--$\sigma$--range  for  the random  clusters.   Whereas the  scalar
Minkowski  functionals  are consistent  with  the  Poisson model,  the
curvature  centroids,   especially  their  $y$--component,   reveal  a
significant deviation from this model.
\label{fig:poissonvgl}}
\end{figure}
-- To  strengthen  our  conclusions  we also  simulated  the  Poisson
clusters  with   a  different   binning,  but  the   results  remained
stable. The  results demonstrate that the  cluster Cl0016+161 exhibits
evidential subclustering in comparison with a Poissonian model.


\section{Conclusions}\label{sec:con}
The examples discussed in this  note support our claim that the method
of  cluster analysis  based  on the  combination  of scalar  Minkowski
functionals  with vectorial  centroids  furnishes a  versatile set  of
order parameters  to sort out essential aspects  of cluster morphology
such as symmetry, clumpiness,  global shape and topology.  Our cluster
morphometry rests  on a  solid mathematical basis  derived from  a few
reasonable  requirements.   In this  sense,  it  allows  for a  unique
morphological description. The construction of patterns from empirical
data   introduces  additional   parameters  which   may   be  employed
advantageously  for scale--specific  diagnosis.  No  tacit statistical
assumptions are involved.
\\ Finally, we note that  these families of morphological measures may
be extended  further to include  tensor--valued Minkowski functionals,
which generalize the concept of inertia tensors.
\\
The  code to  compute the  Minkowski functionals  and  the Querma{\ss}
vectors    is    available    on    request    from    the    authors.

\section*{Acknowledgements}
We thank J. Schmalzing for providing his 3d--code for computing scalar
Minkowski  functionals  and  for  useful  comments,  J.   Colberg  for
providing the GIF-clusters and  M.  Kerscher for valuable discussions.
This work was supported  by the ``Sonderforschungsbereich 375-95 f\"ur
Astro-Teilchenphysik''                  der                  Deutschen
Forschungsgemeinschaft.   T.B.  acknowledges   generous   support  and
hospitality by the National Astronomical Observatory in Tokyo, as well
as hospitality at Tohoku University in Sendai, Japan.

\bibliographystyle{abbrv}


\end{document}